\documentclass{aa}  

\usepackage{graphicx,txfonts,lscape,natbib}
\bibpunct{(}{)}{;}{a}{}{,} % to follow the A&A style

\begin{document}

   \title{A new L5 brown dwarf member of the Hyades cluster with chromospheric activity\thanks{Based on observations made with the Gran Telescopio de Canarias (GTC) installed at the Spanish Observatorio del Roque de los Muchachos of the Instituto de Astrof\'isica de Canarias, on the island of La Palma.}}

%   \subtitle{}

   \author{A.\ P\'erez-Garrido \inst{1}
        \and
        N.\ Lodieu \inst{2,3}
        \and
        R.\ Rebolo \inst{2,3,4}
        }

   \institute{Dpto.\ F\'\i sica Aplicada, Universidad Polit\'ecnica de Cartagena, E-30202  Cartagena, Murcia, Spain\\
       \email{antonio.perez@upct.es}
       \and
       Instituto de Astrof\'isica de Canarias (IAC), Calle V\'ia L\'actea s/n, E-38200 La Laguna, Tenerife, Spain
       \and
       Departamento de Astrof\'isica, Universidad de La Laguna (ULL), E-38206 La Laguna, Tenerife, Spain
       \and
       Consejo Superior de Investigaciones Cient\'ificas, CSIC, Spain
             }

   \date{Received \today{}; accepted (date)}

% \abstract{}{}{}{}{} 
% 5 {} token are mandatory
 
  \abstract
   {}
   {Our aim is to identify brown dwarf members of the nearby  Hyades open star cluster to determine the photometric and spectroscopic properties of brown dwarfs at moderately old ages  and extend the  knowledge  of the substellar mass function of the cluster.
   }
   {We cross-matched the 2MASS and AllWISE public catalogues and measured proper motions to  identify low-mass 
stars and brown dwarf   candidates in an area of  radius eight degrees around the central region of the 
Hyades cluster. We identified objects with photometry and  proper motions consistent with cluster membership. 
For the faintest ($J$\,=\,17.2 mag) most promising astrometric and photometric low-mass candidate 
2MASS\,J04183483$+$2131275\@,  with a membership probability of 94.5\%, we obtained low-resolution (R\,=\,300--1000) and intermediate-resolution (R\,=\,2500)  spectroscopy with the  10.4m Gran Telescopio Canarias.   
   }
   {From the low-resolution spectra we determined a L5.0$\pm$0.5 spectral type, consistent with the  available photometry.
 In the intermediate dispersion spectrum we detected H$\alpha$ in emission (marginally resolved with a 
full width half maximum of $\sim$2.8 \AA{}) and determined a $\log ($L$_{\rm{H}\alpha}$/L$_{\rm bol})=-6.0$ dex. 
From H$\alpha$ we obtained a  radial velocity of 38.0$\pm$2.9 km$\cdot$s$^{-1}$, 
which combined with the proper motion leads to space velocities which are fully consistent with membership 
in the  Hyades cluster. We also report a detection in the H$_2$ band by the UKIDSS Galactic Plane
Survey. Using evolutionary models we determine from the available photometry 
of the object a mass in the range 0.039$-$0.055  M$_{\odot}$. Brown dwarfs with masses below 
0.055 M$_{\odot}$ should fully preserve its initial lithium content, and indeed the spectrum at 6708 \AA{} 
may show a feature consistent with lithium preservation;
however, a higher S/N is needed to confirm this point.  
 }
   {We have identified a new high-probability L5 brown dwarf member of the Hyades cluster. 
This is the first relatively old  L5 brown dwarf with a well-determined age (500$-$700 Myr) and measured 
chromospheric emission.}  

  \authorrunning{P\'erez Garrido et al$.$}
  \titlerunning{A new L5 member of the Hyades}

   \maketitle
%
%________________________________________________________________

%
%%%%%%%%%%%%%%%%%%%%%%%%%%%%%%
%%%%%  Introduction  %%%%%
%%%%%%%%%%%%%%%%%%%%%%%%%%%%%%
%
\section{Introduction}
\label{HyadesL5:intro}

Brown dwarfs are objects with masses below 0.075 $M_{\odot}$ unable to reach high enough temperatures 
in their cores to trigger hydrogen fusion \citep{burrows93,chabrier97}. The absence of
stable hydrogen burning causes the physical properties of brown dwarfs to change drastically
as a function of mass and age. Dynamical determination of masses has been possible for a
growing list of binary brown dwarfs (e.g.\ \citealt{zapatero04b,dupuy12}), but  the 
age is also required to fully constrain evolutionary models of substellar objects 
\citep{baraffe98,siess00,baraffe15,feiden15a}. The characterization of brown dwarf members of open
star clusters of well-known age is of capital importance because we are provided with snapshots of
the physical properties (colours, luminosities, spectral energy distributions) of these objects 
at given ages. Searches in star-forming regions and young clusters have revealed brown 
dwarfs with ages from a few Myrs to a few hundred Myrs \citep[see reviews by][]{bastian10,luhman12b}.

Older star clusters ($>$\,500\,Myr) like the Hyades (Melotte 25, $\alpha_{2000}$\,=\,04$^{\rm h}$26$^{\rm m}$54$^{\rm s}$, 
$\delta_{2000}$\,=\,$+$15$\hbox{$^\circ$}$52$\hbox{$^\prime$}$) offer a  unique opportunity to 
find examples of relatively old brown dwarfs with  well-determined age and metallicity. The Hyades 
has a mean distance of 46.3$\pm$0.3 pc and is the closest open star cluster to the Sun \citep{vanleeuwen09}.
The Hyades exhibit a significant mean proper motion: $\mu=$\,$\sim$\,74--140 mas/yr, 
PA\,=\,90$^{\rm o}$-135$^{\rm o}$ \citep{bryja94}, a tidal radius of $\sim$10 pc, and a core radius 
of 2.5--3.0 pc \citep{perryman98}. Owing to its proximity, the cluster spans a large area over the sky 
(several hundred  square degrees). The age of the Hyades is 625$\pm$50 Myr based on the comparison 
of the observed cluster sequence with model isochrones, although a wider age range  cannot be discarded 
\citep{mermilliod81,eggen98a}. The metallicity of high-mass members appears slightly super-solar, around 
[Fe/H]\,$\sim$\,0.13--0.14 dex \citep{boesgaard90} although a more recent work by \citet{gebran10} 
suggests a mean solar metallicity. 

\citet{hogan08} reported a number of L dwarf candidates in the Hyades cluster based on photometry
and proper motions. They identified 12 objects with colours resembling those of field L dwarfs. 
Recently, \citet{casewell14a} and \citet{lodieu14b} presented spectroscopic follow-up confirming
the cool nature of most of the candidates reported by \cite{hogan08}, although these authors did not
unambiguously determine whether these objects are brown dwarfs or very low-mass stars.
\cite{bouvier08a} discovered the first two T-type dwarfs in the Hyades cluster based on
low-resolution infrared spectra. These cool objects are very likely brown dwarfs. Based on the shape of the mass function, \citet{bouvier08a} argued that $\sim$15 brown dwarfs 
could exist in the present-day Hyades cluster and are yet to be uncovered.
In addition to 2MASSI\,023301.55$+$2470406 \citep{cruz07}, which was proposed as a L0 member of the Hyades \citep{goldman13}, a few other known L/T dwarfs listed with spectra and proper motions in the compendium of
ultracool dwarfs\footnote{see http://spider.ipac.caltech.edu/staff/davy/ARCHIVE/index.shtml} 
could be associated with the Hyades moving group \citep{bannister07,gagne15c} and therefore could be relatively old.

In this paper, we present the finding of a new mid-L dwarf, 2MASS\,J04183483$+$2131275 (hereafter 2M0418$+$21) with photometry, spectroscopy, proper motion, and radial velocity consistent with membership in the Hyades cluster. Furthermore, it presents 
signs of chromospheric activity with a clear detection of H$\alpha$ in emission on 19 December 2015  for a few hours. In Sect.\ \ref{HyadesL5:new_memb} we describe our search for new  very low-mass  members of the Hyades.
In Sect.\ \ref{HyadesL5:spec_obs} we present spectroscopic observations of the most promising candidate found 2M0418$+$21  conducted with the Gran Telescopio de Canarias (GTC).
In Sect.\ \ref{HyadesL5:analysis} we analyse the optical spectra and put our results
in context with other low-mass stars and brown dwarfs.

%
%%%%%%%%%%%%%%%%%%%%%%%%%%%%%%%%%%%%%%%%%%
%%%%% New low-mass Hyades members  %%%%%
%%%%%%%%%%%%%%%%%%%%%%%%%%%%%%%%%%%%%%%%%%
%
\section{New very low-mass proper motion member of the Hyades}
\label{HyadesL5:new_memb}
\subsection{Catalogue cross-match}
\label{HyadesL5:new_memb_Xmatch}

\begin{figure}
\resizebox{\hsize}{!}{\includegraphics[angle=-90]{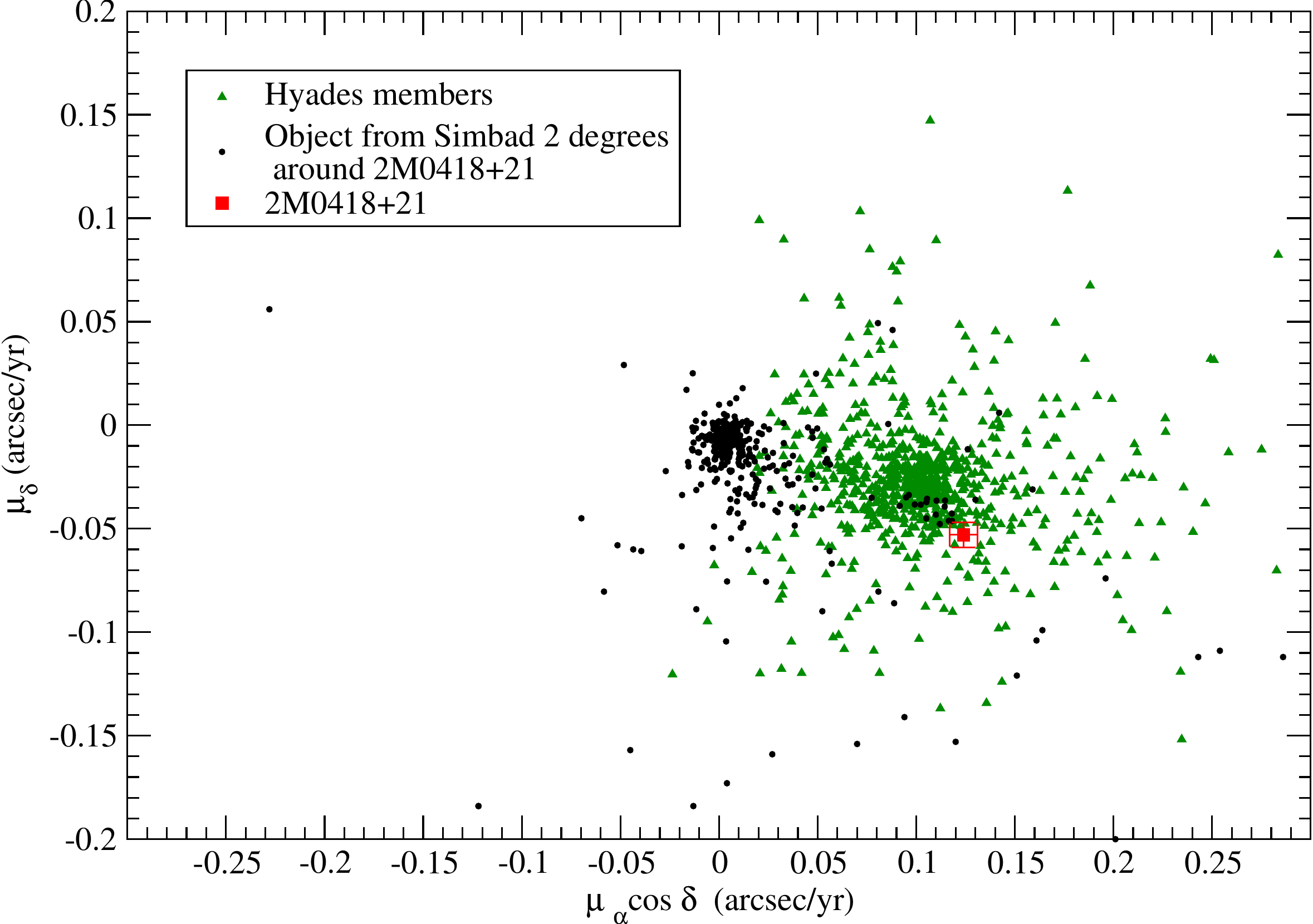}}
\caption{Proper motion diagram of the Hyades cluster. Green triangles correspond to Hyades members 
\citep{goldman13} and the solid square corresponds to 2M0418$+$21\@. Black dots corresponds to objects 
in the Simbad database within a circular radius of 2 degrees around 2M0418$+$21\@.
}
\label{fig_HyadesL5:PM}
\end{figure}

We cross-matched the Two Micron All Sky Survey \citep[2MASS;][]{cutri03,skrutskie06} point 
source catalogue and the AllWISE mid-infrared catalogue built upon the Wide-field Infrared Survey Explorer 
mission \citep[WISE;][]{wright10} to identify new low-mass stars and substellar members in the Hyades cluster.
 Details of this cross-match will be presented in a future paper 
(P\'erez-Garrido et al.\ 2016, in prep.).
Given the temporal baseline between the two surveys ($\sim$11--14 yr) we conducted an astrometric
search to identify objects with mean displacements larger than 100 mas/yr. In our search, we 
looked for pairs of uncorrelated objects in 2MASS and WISE in a Hyades region of radius 8 degrees 
  centred at 04$^{\rm h}$26$^{\rm m}$54$^{\rm s}$, $+$15$^{\rm o}$52$^\prime$.
Uncorrelated objects  are those which are present in one catalogue 
with no counterpart in the other database within a correlation radius of one arcsec. In a subsequent 
step, we selected objects with apparent proper motions measured between both catalogues of 
$70<\mu_{\alpha}\cos \delta<130$ mas/yr in right ascension and $-60<\mu_{\delta}<0$ mas/yr
in declination.
These pairs represent candidates with proper motions comparable to those of 
Hyades members (Fig.\ \ref{fig_HyadesL5:PM}). 
This correlation returned 130 objects with proper motions and near-infrared photometry 
compatible with cluster membership (Fig.\ \ref{fig_HyadesL5:PM}; Fig.\ \ref{fig_HyadesL5:MW2}). 
We recovered several L dwarf candidates \citep{hogan08} with spectroscopic follow-up
\citep{casewell14a,lodieu14b}.  A subset of those 130 candidates found in our search 
also have  optical photometry in the Sloan Sky Digital Survey \citep{york00}.
Remarkably, we identified a new object whose colours measured by  2MASS, WISE, and the Sloan Digital Sky Survey
\citep[SDSS;][]{york00}  were indicative of 
a mid-L dwarf (Fig.\ \ref{fig_HyadesL5:MW2}; Table \ref{tab_HyadesL5:phot}) that is cooler and 
significantly fainter than other previously reported  Hyades L dwarfs. 
In Sect.\ \ref{HyadesL5:spec_obs} we present the optical 
spectral classification of this object.

\subsection{Proper motion computation}
\label{HyadesL5:new_memb_PM}
The proper motion of 2M0418$+$21 was obtained using the 2MASS, AllWISE, and UKIDSS GCS catalogues and simply calculated as $\mu_\alpha=\Delta \alpha\cos\delta/\Delta t$ and $\mu_\delta=\Delta \delta/\Delta t$, 
where $\Delta \alpha$ and $\Delta \delta$ are the differences between RA and Dec in any two catalogues.
We note that we used the set of coordinates not corrected for proper motion in the AllWISE catalogue.
In order to assess the error in our measurement for each pair of catalogues we determined the proper motions
of all objects placed 30 arcmin around 2M0418$+$21\@. We estimate the error, $\sigma$, from the full width half maximum of 
each bell-shaped proper motion histogram (2.3 times $\sigma$). We discarded the proper motion
measurements from the AllWISE vs UKIDSS GCS pairs owing to the short baseline between the two observations as it yielded large error bars 
($\sim$60 mas/yr). For the other two pairs of catalogues we obtained errors of $\sim$8 mas/yr and the initial resulting 2M0418$+$21 proper motions were (129,$-$53) mas/yr and (133,$-$44) mas/yr for 2MASS-AllWISE and 
2MASS-UKIDSS GCS, respectively.  However, for the Hyades objects a correct estimation of the proper motion requires taking into account  parallactic motion. We  removed this parallactic effect using the Earth's barycentre coordinates for each catalogue epoch, assuming a distance of 48.8 pc.  We then performed  a linear regression with the new coordinates from 2MASS, UKIDSS, and WISE AllSky (we did not use AllWISE since its astrometry is an average spanning several months), and the  resulting fit yielded a final proper motion of (124$\pm$7,-53$\pm$6) mas/yr   for our object.

%
%%%%%%%%%%%%%%%%%%%%%%%%%%%%%%
%%%%% Figure: PM  %%%%%
%%%%%%%%%%%%%%%%%%%%%%%%%%%%%%
%

\subsection{Hyades membership}
\label{HyadesL5:new_memb_CP}
All members of the Hyades cluster seem to move towards the cluster convergent point (CP) located at $\alpha_{\rm CP}=$6$^{\rm h}$29.48$^{\rm m}$, $\delta_{\rm CP}$=6$^{\rm o}$53$^\prime$.4 \citep{madsen02}. The angle $\theta$ from this object to the convergent point (measured as the angle between the line pointing to the north and the line to the CP) can be calculated as in  \cite{hogan08}. This angle should be compared to the proper motion angle, $\theta_{\rm PM}$. Values for these angles with differences below 12$^{\rm o}$  would indicate that our object is moving towards the Hyades CP. In our case we get $\theta=$110$^{\rm o}$ and $\theta_{\rm PM}$=113$^{\rm o}$. We also calculate a proper motion distance of 41.7 pc using Eq.\ (3) of \cite{hogan08}. Both estimations are in  good agreement with what we expect for Hyades members \citep{vanleeuwen07}.

Furthermore, we use the maximum likelihood method described in \citet{sanders71} to assess the Hyades membership probability of 2M0418$+$21. This method assumes two overlapping normal bivariate frequency functions, one for field stars and one for the cluster. In our implementation we input the known proper motion of the Hyades cluster, $\mu_\alpha=115$ and $\mu_\delta=-40$ (mas/yr), and let the algorithm estimate the mean proper motion of field objects. We considered all point sources within a radius of 30 arcmin around 2M0418$+$21\@,  finding a membership probability of 
94.5\% for  2M0418$+$21. Therefore, this object can be considered  from both photometry and proper motion as a member of the Hyades cluster.

%
%%%%%%%%%%%%%%%%%%%%%%%%%%%%%%%%%%%%%%%%%
%%%%% Figure: finding chart  %%%%%
%%%%%%%%%%%%%%%%%%%%%%%%%%%%%%%%%%%%%%%%%
%
\begin{figure}
 \centering
  \includegraphics[width=\linewidth, angle=0]{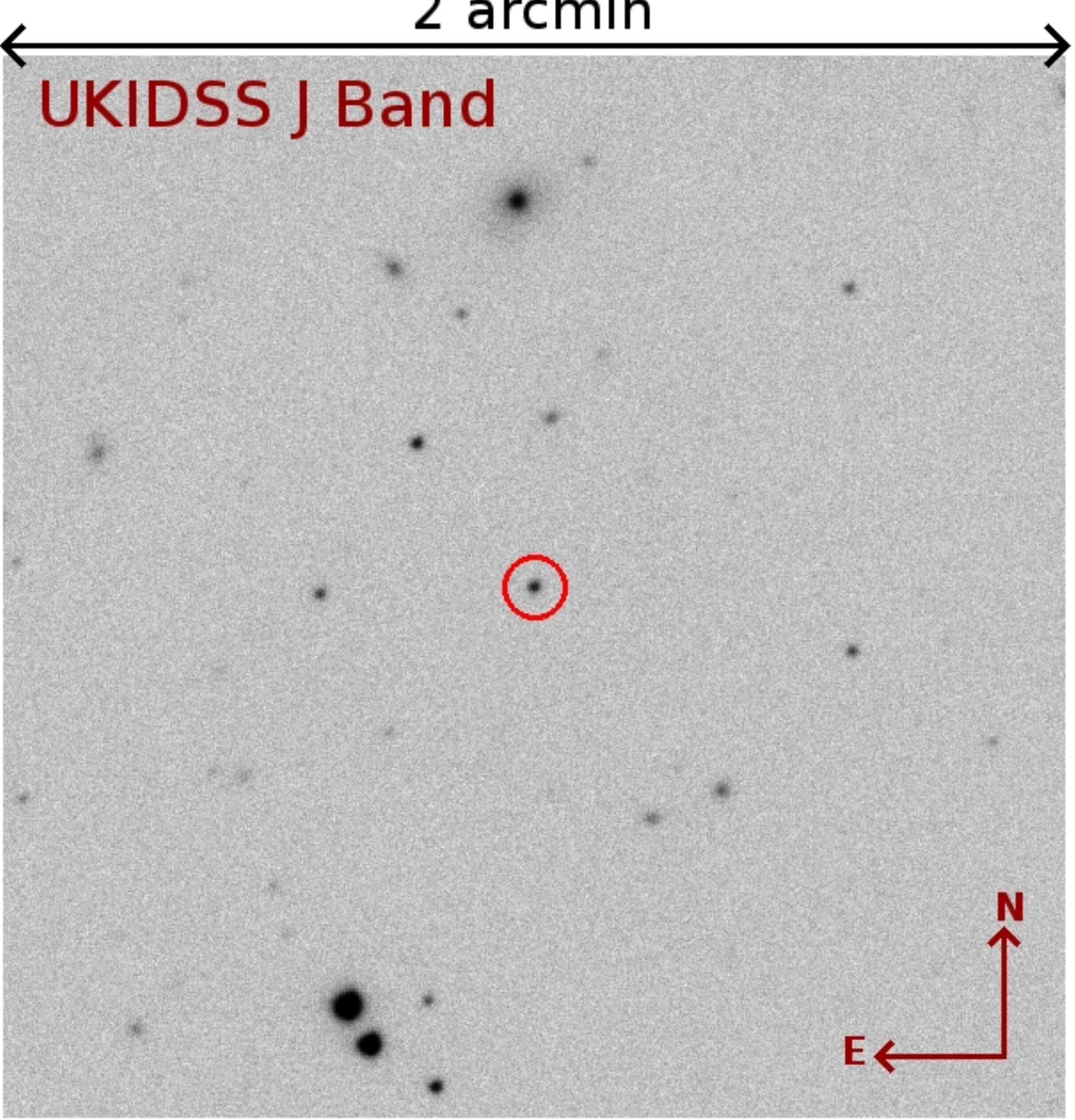}
   \caption{Finding chart of 2M0418$+$21 in J band from UKIDSS Galactic Cluster Survey.}
   \label{fig_HyadesL5:chart}
\end{figure}
%

%
%%%%%%%%%%%%%%%%%%%%%%%%%%%%%%%%%%%%%%
%%%%% Figure: (w1-w2) vs (J-K)  %%%%%
%%%%%%%%%%%%%%%%%%%%%%%%%%%%%%%%%%%%%%
%
\begin{figure}
  \includegraphics[width=.9\linewidth, angle=-90]{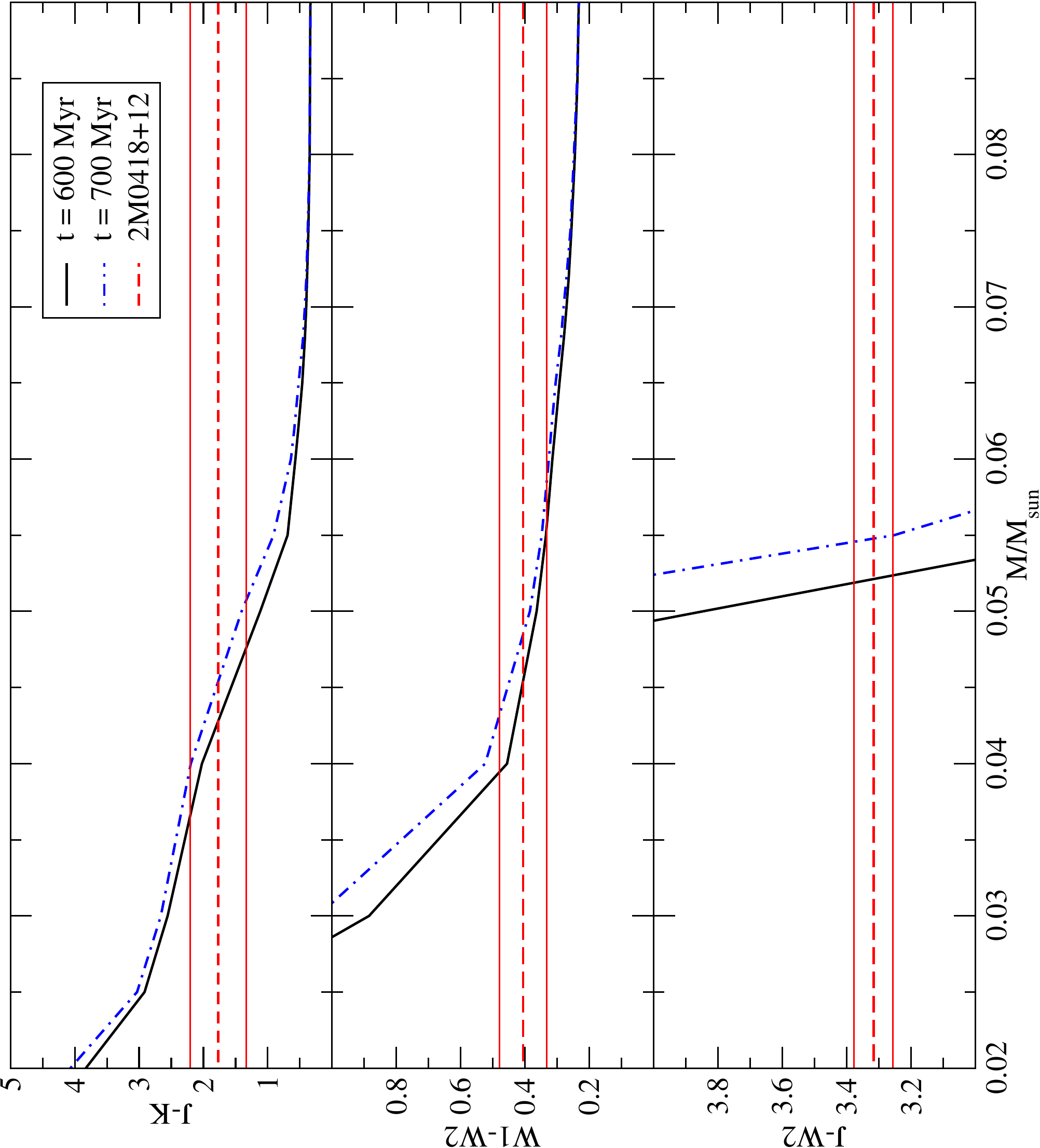}
\caption{$J-K$, $W1-W2$, and $J-W2$ colour as a function of mass from the AMES-Dusty evolutionary model.
The dashed red line with the bottom and top red line represent the $W1-W2$ (AllWISE) and $J-K$ (UKIDSS) colours
of 2M0418$+$21 with their error bars.
}
\label{fig_HyadesL5:M_W1-W2}
\end{figure}
%

%
%%%%%%%%%%%%%%%%%%%%%%%%%%%%%%%%%%%%%
%%%%% Figure: CMD (w1-w2,Mw2)  %%%%%
%%%%%%%%%%%%%%%%%%%%%%%%%%%%%%%%%%%%%
%
\begin{figure}
  \includegraphics[width=\linewidth, angle=0]{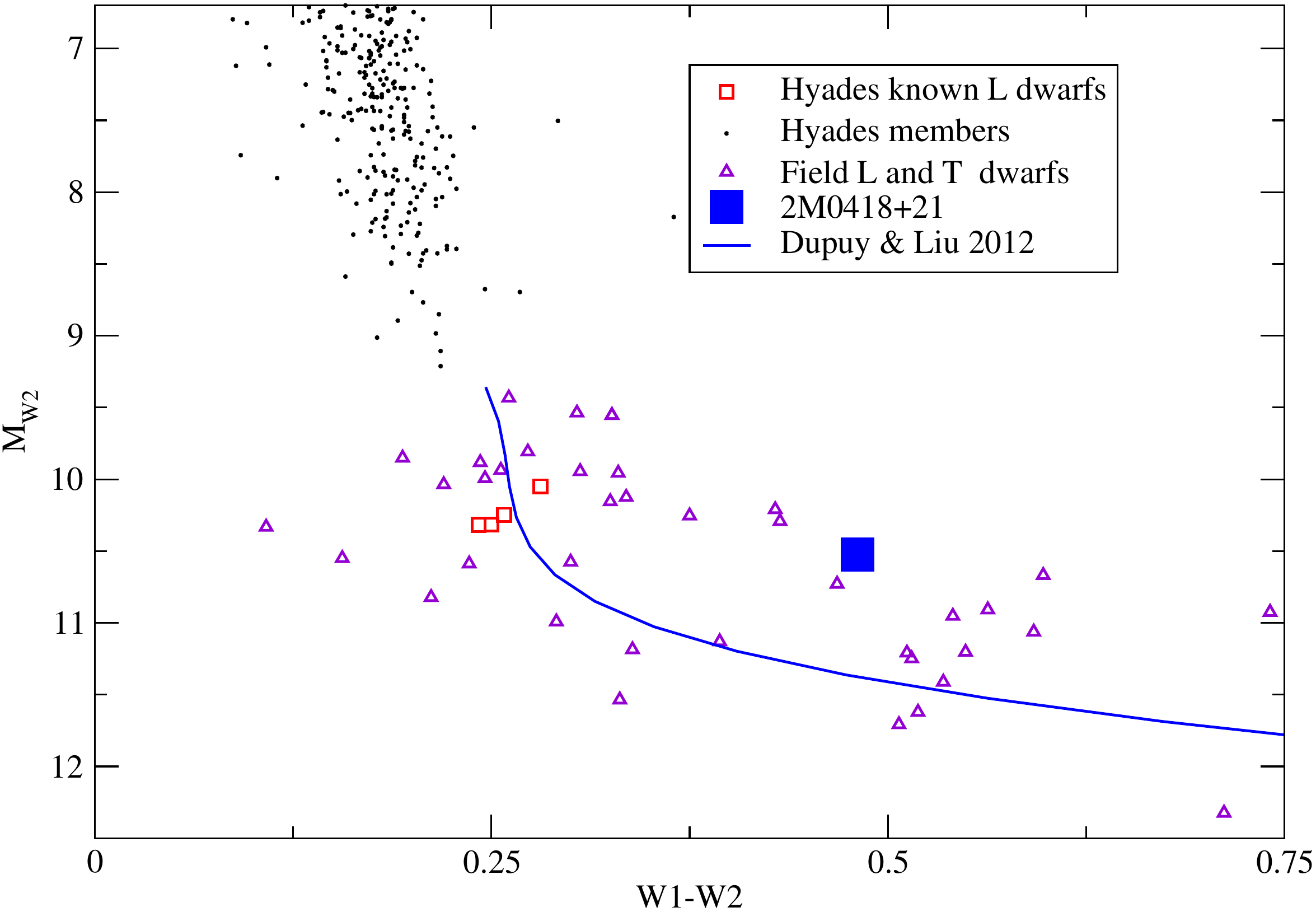}
%\resizebox{.95\hsize}{!}{\includegraphics{MW2_W1-W2.eps}}
\caption{($W1-W2$,M$_{W2}$) colour-magnitude diagram depicting known L dwarfs in the Hyades 
\citep[red squares;][]{hogan08,casewell14a,lodieu14b}, Hyades high-mass and low-mass members 
\citep[black dots;][]{goldman13}, the mean sequence of field L and T dwarfs shifted to 46.3 pc 
\citep[purple line;][]{dupuy12}, and our target 2M0418$+$21 (solid square; this work).
}
\label{fig_HyadesL5:MW2}
\end{figure}

\subsection{Additional photometry}
\label{HyadesL5:new_memb_photometry}

In addition to the $JHK_{s}$ near-infrared photometry from 2MASS and mid-infrared data from WISE, 
2M0418$+$21 is also covered by the Sloan Sky Digital Survey Data Release 9 \citep{york00,ahn12a} and 
the UKIRT Infrared Deep Sky Survey \citep[UKIDSS;][]{lawrence07} Galactic Clusters Survey and 
Galactic Plane Surveys. 2M0418$+$21 is detected in the Sloan $z$ band only and in various filters
from UKIDSS with two distinct epochs (1 December 2005 for $K1$ and 29 November 2005 for $K$).
We summarize the photometry available for 2M0418$+$21 in Table \ref{tab_HyadesL5:phot}
and display a $J$-band finding chart from UKIDSS in Fig.\ \ref{fig_HyadesL5:chart}.

\begin{figure*}
\resizebox{\hsize}{!}{\includegraphics{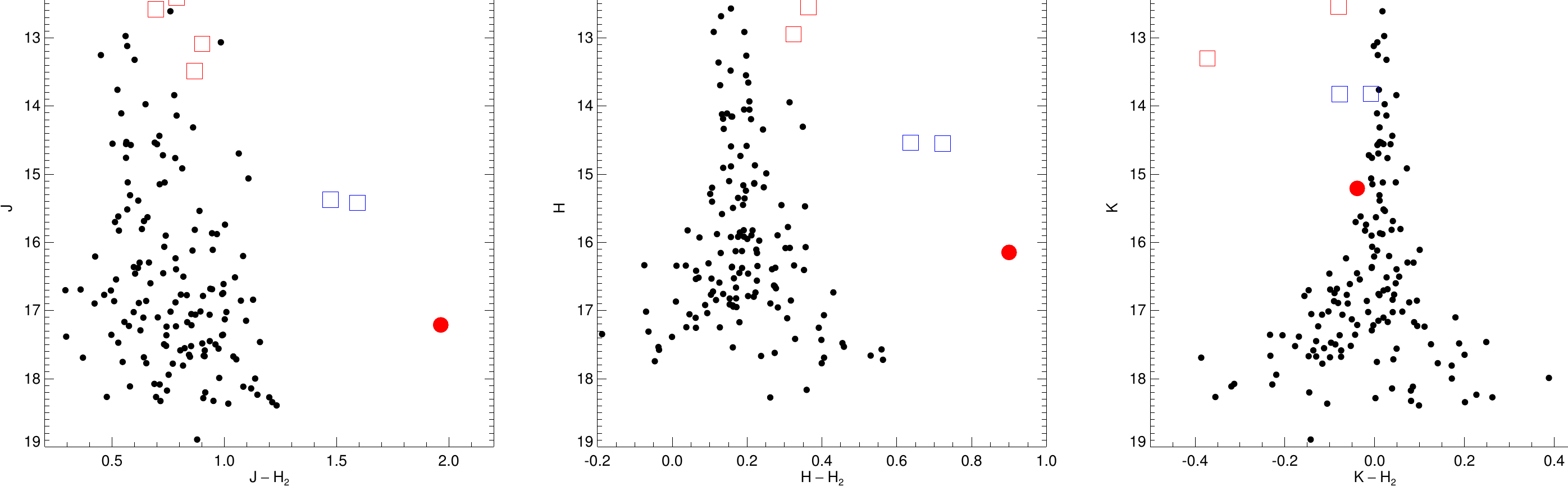}}
\caption{{\it{Left:}} ($J-H_{2}$,$J$) colour-magnitude for all point sources within 5
arcmin of 2M0418$+$21 highlighted by a large red dot.
{\it{Middle:}} Same but for ($H-H_{2}$,$H$).
{\it{Right:}} Same but for ($K-H_{2}$,$K$). Hyades members from \citet{goldman13} and field L/T dwarfs with H$_{2}$ photometry from the UKIDSS GPS are plotted as red and blue open squares, respectively.}
\label{fig_HyadesL5:H2_CMDs}
\end{figure*}

We checked the $H_{2}$(1--0)-band image obtained by the UKIDSS Galactic Plane Survey on
29 November 2005\@. The DR10 catalogue gives a magnitude of 15.247$\pm$0.029 mag with an
ellipticity below 0.06 and roundness parameters indicating a point source.
However, 2M0418$+$21 stands out in the ($J-H_{2}$,$J$) and ($H-H_{2}$,$H$) colour-magnitude
diagrams in comparison to other point-like objects within five arcmin (Fig.\ \ref{fig_HyadesL5:H2_CMDs}).
Its $J-H_{2}$ and $H-H_{2}$ colours are redder than any other point source in the field
(Fig.\ \ref{fig_HyadesL5:H2_CMDs}).
However, its location in the ($K-H_{2}$,$K$) colour-magnitude diagram fits  the main 
sequence of point sources well (right side plot in Fig.\ \ref{fig_HyadesL5:H2_CMDs}).
The $H_{2}$(1--0) is centred at 2.12 microns with a 50\% cut-off of 0.1 microns, and is
usually used to probe the presence of molecular emission from circumstellar disks. We do
not see any obvious presence of a jet in the 5 arcmin  image.
We have checked the Herschel and ALMA archives, but the search did not return any public image at the time of writing so we cannot  discuss further the presence or absence of a disk around 2M0418$+$21\@.

We added two samples of objects in Fig.\ \ref{fig_HyadesL5:H2_CMDs} to place the $H_{2}$ photometry
of 2M0418$+$21 in context. We cross-matched the full list of Hyades members from \citet{goldman13}
with the UKIDSS GPS DR10 and found 79 sources with $H_{2}$ photometry (red open squares in
Fig.\ \ref{fig_HyadesL5:H2_CMDs}). However, only a handful have magnitudes fainter than 12 mag.
Only one source has an $H_{2}$ magnitude fainter than 13 mag, safely below the saturation limits of 
the UKIDSS shallow surveys \citep{lodieu07a}. We also included two L dwarfs covered by the
GPS among all field L/T dwarfs listed in the compendium of ultracool 
dwarfs\footnote{see http://spider.ipac.caltech.edu/staff/davy/ARCHIVE/index.shtml}.
One object is 2MASSW\,J0326137$+$295015 classified in the optical as a L3.5 dwarf \citep{kirkpatrick99}
with a trigonometric distance of 32.2$^{+1.49}_{-1.64}$ pc and a mean proper motion of 69.4$\pm$0.8 
mas/yr with a position angle of 344.3$\pm$0.7 degrees \citep{dahn02}. The proper motion
of this source is lower than the average motion of Hyades members with a discrepant position
angle and it lies closer than the Hyades, arguing against its membership in the Hyades.
The other source is 2MASSI\,J0409095$+$210439 \citep[L3.0;][]{kirkpatrick00}. This object lies
close to the centre of the Hyades cluster and has a proper motion of 
101$\pm$15,$-$148$\pm$12 mas/yr \citep{casewell08a} and a spectrophotometric
distance of 35.9--39.8 pc, which makes it a potential L dwarf member of the Hyades cluster.
We observe a possible trend of redder $J-H_{2}$ and $H-H_{2}$ colours with later spectral types
in the L dwarf regime.

\subsection{Mass estimate}
\label{HyadesL5:new_memb_mass}

In Fig.\ \ref{fig_HyadesL5:M_W1-W2} we compare the colours of 2M0418$+$21 with predictions for 
brown dwarfs of various masses at the age of the Hyades using the evolutionary models of
the Lyon group \citep{allard01,chabrier00c}. Each colour gives a slightly different mass estimation: 0.044$^{+0.005}_{-0.005}$ M$_\odot$, 0.046$^{+0.010}_{-0.005}$ M$_\odot$, and $0.054^{+0.001}_{-0.001}$  M$_{\odot}$ for $J-K$, $W1-W2$, and $J-W2$, respectively.
We infer a likely mass between 0.039--0.055 M$_{\odot}$ with an average value of 0.048$^{+0.007}_{-0.009}$  M$_{\odot}$  for 2M0418$+$21. To summarize, we have identified a highly probable brown dwarf member of the Hyades cluster.

%
%%%%%%%%%%%%%%%%%%%%%%%%%%%%%%%%%%%%%%%
%%%%% Table: Photometry target  %%%%%
%%%%%%%%%%%%%%%%%%%%%%%%%%%%%%%%%%%%%%%
%
\begin{table}
\caption{Compilation of the photometry for 2M0418$+$21.}

\label{tab_HyadesL5:phot}
\centering
\begin{tabular}{c c}
\hline
\hline
R.A.\ (2MASS)  &     04:18:34.83   \\
dec (2MASS)   &  $+$21:31:27.5    \\
$z$ (SDSS DR9) & 20.338$\pm$0.116 mag \\
$J$ (2MASS) & 17.152$\pm$0.239 mag \\   
$H$ (2MASS) & 16.265$\pm$0.211 mag \\
$K_{s}$ (2MASS) & 15.381$\pm$0.198 mag \\
$J$ (UKIDSS GPS DR10) & 17.211$\pm$0.016 mag \\
$H$ (UKIDSS GPS DR10) & 16.147$\pm$0.011 mag \\
$K$ (UKIDSS GPS DR10) & 15.208$\pm$0.013 mag \\
$H_{2}$ (UKIDSS GPS DR10) & 15.247$\pm$0.029 mag \\
$K1$ (UKIDSS GCS DR10) & 15.230$\pm$0.015 mag \\
$W1$ (AllWISE) & 14.300$\pm$0.029 mag \\
$W2$ (AllWISE) & 13.894$\pm$0.045 mag \\
$\mu_{\alpha}\cos\delta$ &  124\,$\pm$\,7 mas/yr \\
$\mu_{\delta}$ &  $-$53\,$\pm$\,6 mas/yr \\
\hline
%inserts single line
\end{tabular}
\tablefoot{Photometry in the optical, near-infrared,  and mid-infrared from SDSS DR9 
\citep{york00,adelman_mccarthy12}, 2MASS Point Source catalogue \citep{cutri03,skrutskie06}, UKIDSS Galactic Plane Survey and Galactic 
Clusters Survey \citep{lawrence07,lucas08}, and AllWISE \citep{wright10}, as well as the mean
proper motion determined in this paper. There is no reliable detection in the SDSS $i$ filter and 
the WISE $w3$ band.}
\end{table}

%
%%%%%%%%%%%%%%%%%%%%%%%%%%%%%%%%%%%%%%
%%%%% Figure: H2 CMDs  %%%%%
%%%%%%%%%%%%%%%%%%%%%%%%%%%%%%%%%%%%%%
%

%

%
%%%%%%%%%%%%%%%%%%%%%%%%%%%%%%
%%%%% Observations  %%%%%
%%%%%%%%%%%%%%%%%%%%%%%%%%%%%%
%
\section{Spectroscopic observations}
\label{HyadesL5:spec_obs}

We collected several optical spectra with three different gratings installed on the OSIRIS
\citep[Optical System for Imaging and low-intermediate Resolution Integrated Spectroscopy;][]{cepa00}
spectrograph on the 10.4m GTC in La Palma to characterize the target.
OSIRIS is equipped with two 2048 $\times$ 4096 Marconi CCD42-82 detectors offering a
field of view of approximately 7$\times$7 arcmin$^2$ with a binned pixel scale of 0.25 arcsec.

\subsection{Low-resolution optical spectroscopy}
\label{HyadesL5:analysis_LowRes}

We obtained two low-resolution optical spectra of 2M0418$+$21 with the R300R grating on GTC/OSIRIS
on the nights of 25 and 26 January 2015 under filler program GTC51--14B (PI Lodieu).
We used an on-source integration of 1800 sec for both spectra with the following configuration:
slit of 1 arcsec and 2$\times$2 binning. The observations were obtained under a seeing of 1.3 arcsec,
with grey sky and clear conditions.

We collected two new optical spectra of 2M0418$+$21 with the R1000R grating on GTC/OSIRIS
as part of filler program GTC38--15A (PI Lodieu).
We used the same configuration as above, except for the grating which offers a resolution
that is higher by approximately a factor of 3\@. We set the on-source integration to 1800 sec, 
repeated twice with a shift of 10 arcsec (equivalent to 40 pixels) along the slit.
The seeing was poor, around 1.8 arcsec, but the sky was clear and the moon was within 7 days
of new moon.

\begin{figure}
  \centering
  \includegraphics[width=\linewidth, angle=0]{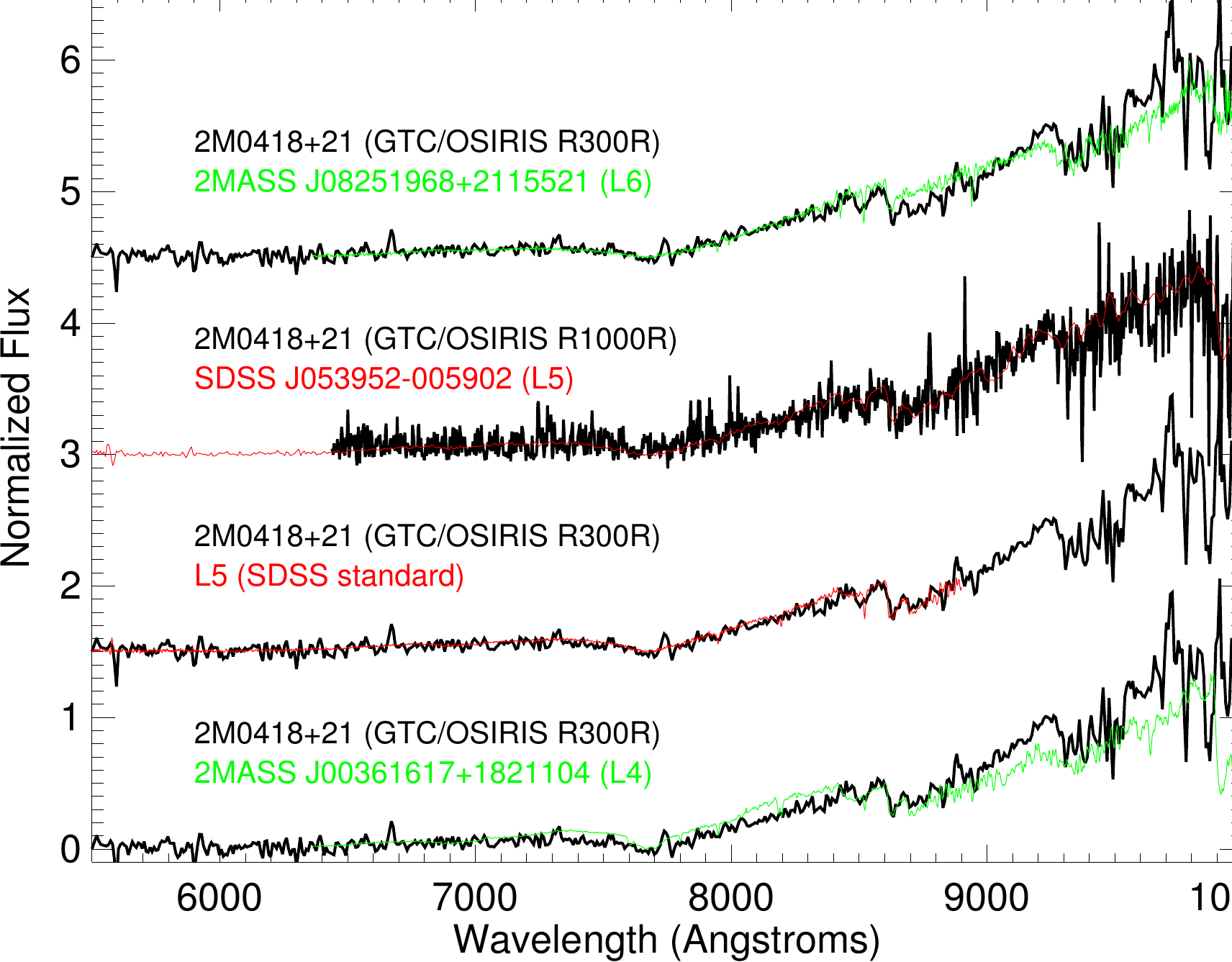}
   \caption{Low-resolution optical spectra of 2M0418 obtained with the R300R (three spectra)
and R1000R (third from bottom) gratings on GTC/OSIRIS (black lines). The continuum of the spectra
are offset vertically by 1.5 for visualization purposes. From top to bottom: We overplotted
the spectra of 2MASS\,J00361617$+$1821104 (L4; green),
the Sloan dwarf template (L5; red), the spectrum of SDSS\,J053952$-$005902 (L5; red),
and 2MASS\,J08251968$+$2115521 (L6; green). See text for details.}
\label{fig_HyadesL5:plot_spec_LowRes}
\end{figure}

We reduced the optical spectra under the IRAF environment \citep{tody86,tody93} in a standard
manner. First, we median-combined the bias and flat fields taken during the afternoon. We
subtracted the mean bias from the raw spectrum of the target and divided by the normalized flat 
field. We extracted the 1D spectrum by optimally choosing 
background level. We calibrated the 1D spectra with the response function derived from the 
spectrophotometric standard star Ross\,640 \citep[DZ5;][]{harrington80,monet03,cutri03,lepine05d,sion09}.
The final spectra of 2M0418$+$21, normalized at 7500 \AA{}, are displayed in 
Figure \ref{fig_HyadesL5:plot_spec_LowRes} along with known L dwarf spectra templates
from the literature \citep{schmidt10b,fan00,leggett02,geballe02}.

%
%%%%%%%%%%%%%%%%%%%%%%%%%%%%%%%%%%%%%%%%%%%%%%%%
%%%%% Figure: Low-resolution spectrum %%%%%
%%%%%%%%%%%%%%%%%%%%%%%%%%%%%%%%%%%%%%%%%%%%%%%%
%
% Figure made with IDL program Hyades_dL_spectra.pro
%
\subsection{Medium-resolution optical spectroscopy}
\label{HyadesL5:analysis_MedRes}

Following the analysis of the low-resolution spectra, we requested a dedicated program to 
take optical spectra with the highest resolution grating available (R2500R)  on GTC/OSIRIS, 
offering a resolution of $\sim$2400 at 6600 \AA{} with a slit of 1 arcsec.
We secured five groups of two spectra of 30\,min on-source each shifted along the slit by 
10 arcsec, yielding a total on-source exposure of 5h. The observations were taken on the 
night of 19 December 2015 under a variable seeing of 1.3--1.5 arcsec, spectroscopic conditions, 
and the moon illuminated at 65\% located at 50 degrees from our target.

\begin{figure}
  \centering
  \includegraphics[width=\linewidth, angle=0]{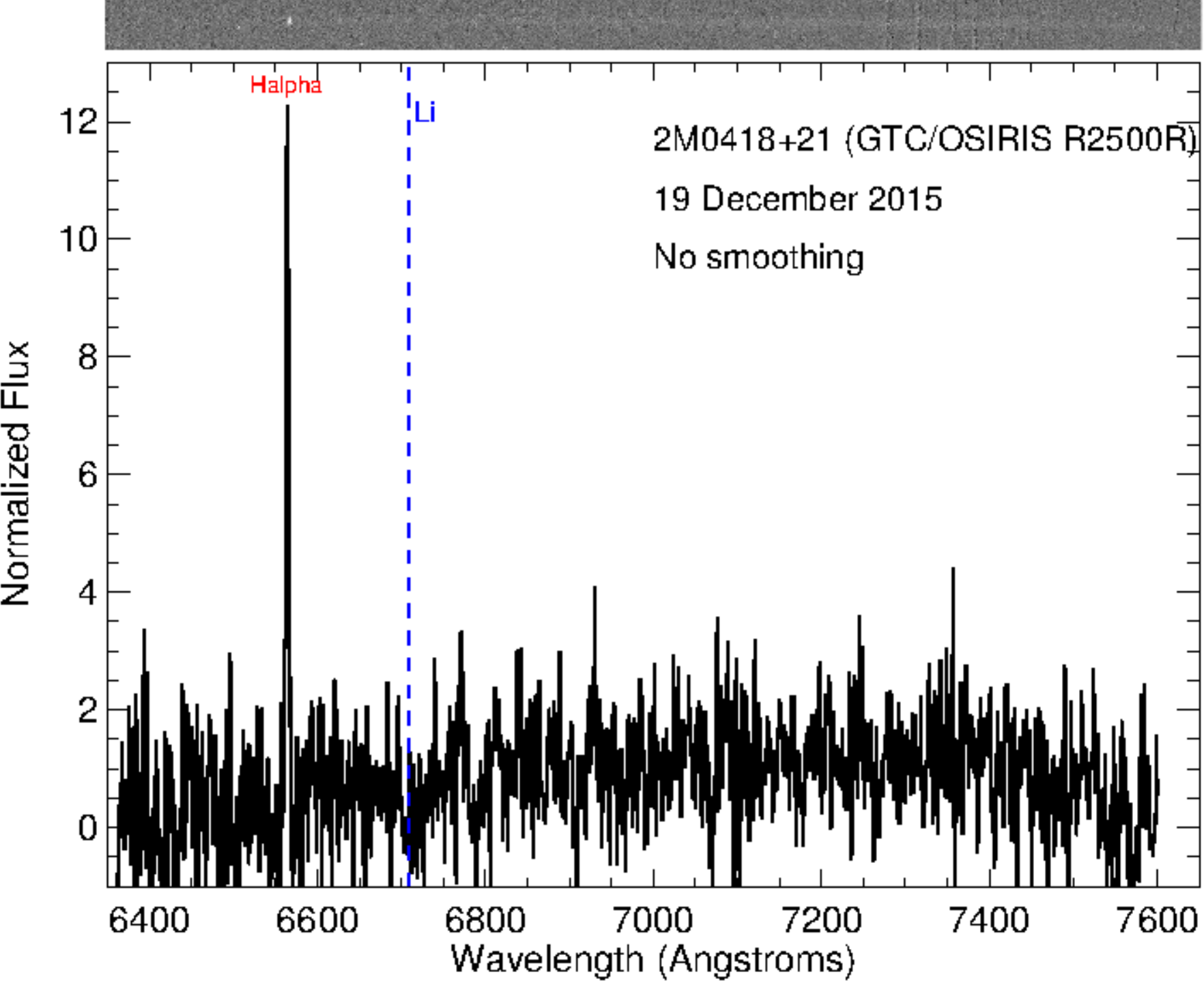}
   \caption{{\it{Bottom:}} Combined 1D optical spectrum of 2M0418$+$21 taken with 
the R2500R grating on GTC/OSIRIS on 19 December 2015\@. No smoothing was applied.
{\it{Top:}} Final combined 2D image of the R2500R spectrum of 2M0418$+$21\@. The white 
dot on the left-hand side corresponds to the H$\alpha$ emission. The faint continuum
is barely seen, but is present.
}
   \label{fig_HyadesL5:plot_spec_MedRes}
\end{figure}

We reduced the R2500R spectra under the IRAF environment \citep{tody86,tody93} taking advantage of
the fact that all spectra were obtained consecutively on the same night. We subtracted each
pair of optical spectra to remove the contribution of the sky. Then, we averaged the five
spectra at the nominal position (``first frame'') and the shifted position (``second frame'') 
rejecting the lowest and highest values to create a combined spectrum at each position. Later, 
we combined the two frames by shifting the second frame to the first one. The final 2D image is 
shown at the top of Fig.\ \ref{fig_HyadesL5:plot_spec_MedRes}, where the H$\alpha$ emission
line stands out as a white dot with a signal-to-noise of 10 on top of a very faint continuum 
with a signal-to-noise of $\sim$2 per spectral resolution element of 2.6 \AA.

We optimally extracted   the spectrum by selecting the nominal position fitting the
background without cosmic rays rejection. 
We calibrated our spectrum in wavelength by fitting a 1D polynomial to 20 arc lines
with a dispersion of 1.03 \AA{}/pixel and a rms of 0.05 \AA{}, corresponding to radial velocity 
precisions of $\sim$2.3 km/s. We calibrated this spectrum with the response 
function of the detector$+$grating derived from the spectrophotometric standard star Feige\,110 
\citep{hog00,vanleeuwen07,drilling13}. The final 1D spectrum of 2M0418$+$21 is displayed in
Fig.\ \ref{fig_HyadesL5:plot_spec_MedRes}. The emission line corresponds to H$\alpha$ at
6563 \AA{}.  There is no evidence of H$\alpha$ emission in the spectrum gathered with the R300R grism; however,  data from the R1000R grism show signs of H$\alpha$ in emission with a 
pseudo-equivalent width (pEW) between  4--10 \AA{}, although its low signal-to-noise ratio does not make 
it a reliable detection. 

%
%%%%%%%%%%%%%%%%%%%%%%%%%%%%%%%%%%%%%%%%%%%%%%%%
%%%%% Figure: medium-resolution spectrum %%%%%
%%%%%%%%%%%%%%%%%%%%%%%%%%%%%%%%%%%%%%%%%%%%%%%%
%
% Figure made with IDL program Hyades_dL_spectra.pro
%
%

%
%%%%%%%%%%%%%%%%%%%%%%%%%%%%%%
%%%%% Analysis  %%%%%
%%%%%%%%%%%%%%%%%%%%%%%%%%%%%%
%
\section{Analysis}
\label{HyadesL5:analysis}
\subsection{Spectral classification and spectroscopic distance}
\label{HyadesL5:analysis_SpT}

The overall appearance of the 2M0418$+$21 low-resolution spectrum  displayed in 
Fig.\ \ref{fig_HyadesL5:plot_spec_LowRes} suggests a mid-L dwarf.
To derive a spectral type with better accuracy, we compared our GTC/OSIRIS spectrum with the
Sloan L dwarf template publicly available for each subclass \citep{schmidt10b,schmidt14a}.
We find the L5 SDSS template (red line) provides the best fit to our spectrum (black line)
up to 9000 \AA{} (Fig.\ \ref{fig_HyadesL5:plot_spec_LowRes}). 
We  also compared our optical spectrum to a set of spectral templates listed on
Sandy Leggett's webpage\footnote{http://staff.gemini.edu/$\sim$sleggett/LTdata.html}
and discovered by the Sloan team \citep{knapp04,golimowski04a,chiu06}, including
2MASS\,J00361617$+$1821104 \citep[L4; ][]{reid00,kirkpatrick00,knapp04},
LHS\,102B \citep[L4.5--L5 (binary);][]{goldman99,henry06,kirkpatrick01},
SDSS05395199$-$0059020 (L5),
SDSS13262982$-$0038315 \citep[L5.5;][]{fan00,vrba04}
and 2MASS08251968$+$2115521 \citep[L6;][]{kirkpatrick00,dahn02}.
We find the best match for the L5 dwarf up to 10000 \AA{}, as in the case of the SDSS templates.
The L4 and L6 dwarfs are bluer and redder beyond 8800 \AA{}, respectively 
(Fig.\ \ref{fig_HyadesL5:plot_spec_LowRes}). 
The L6 has a much straighter redward spectrum than our target, marked by the presence
of the FeH bands at $\sim$8700 \AA{} and the H$_{2}$O band at $\sim$9300 \AA{}.
As a consequence, we derive an uncertainty of half a subclass for our optical spectral
classification by comparison with the mid-L dwarf standards.

Assuming the spectral type vs absolute magnitude relation for L5 dwarfs from \citet{dupuy12}
in all five passbands available for 2M0418$+$21, we derive spectrophotometric distances
of 57.8, 53.1, 50.1, 42.3, and 40.6 pc in $J$, $H$, $K$, $w1$, and $w2$, respectively,
with typical errors in the 5--10 pc range. These error bars take into account the uncertainty of 
0.5 class in the spectral type. We observe a trend towards closer distances with redder wavelengths.
Averaging all these measurements with equal
weights because we assume that the polynomial fits in each filter represent independent measurements, 
we infer a mean distance of 48.8$\pm$4.0 pc. The error on the distance is rms on the distance divided by the square 
root of 5 (the number of measurements). We assume that 2M0418$+$21 is single. Our mean value is 
close to the accepted distance of the Hyades cluster from Hipparcos \citep[46.3$\pm$0.3 pc;][]{vanleeuwen09}.

\subsection{H$\alpha$ emission}
\label{HyadesL5:analysis_Halpha}

We detected a clear emission line present in each individual spectrum at the nominal position
of H$\alpha$ (6562.8 \AA{}) over the full time of observations, from UT=22h to UT=1h40 
(Fig.\ \ref{fig_HyadesL5:plot_spec_MedRes}). This is the first detection of H$\alpha$ in emission 
in a mid-L dwarf classified as a high-probability photometric, astrometric, and spectroscopic 
candidate member of the Hyades cluster. However, comparable levels of activity have been reported
in field L dwarfs \citep{schmidt15,pineda16}.

We measured the ratio of H$\alpha$ to bolometric luminosity for our target following the method
outlined in \citet{burgasser11b}. We calibrated the flux of 2M0418$+$21 with the spectrophotometric 
standard star Feige\,110 observed with GTC/OSIRIS on the same night. We considered the $R$-band magnitude 
of Feige\,110 \citep[11.70 mag;][]{landolt07} because it contains the H$\alpha$ line. In this passband our 
target is $\sim$30,000 times fainter than Feige\,110\@. We assumed a bolometric correction in the $K$ band of 
3.3 for a L5 dwarf \citep{golimowski04a} and assumed a solar bolometric luminosity of
M$_{\rm bol}$\,=\,4.74\@. We used the distance of 48.8 pc for 2M0418$+$21 (see previous section).
We infer a value of $\log_{10}$(L$_{{\rm H}\alpha}$/L$_{\rm bol}$) = $-$6.0 dex, consistent with the 
drop in activity seen from late-M to mid-L dwarfs (see e.g.\ Figure 5 of \citealt{berger10}).
Owing to the cluster age we can discard accretion as the origin of this emission and postulate 
chromospheric activity as its cause.

2M0418$+$21 represents an important addition to the sample of L5 dwarfs with detected
H$\alpha$ emission and $\log_{10}$(L$_{{\rm H}\alpha}$/L$_{\rm bol}$) measurements because its
age is well constrained to 625$\pm$50 Myr. We collected a large number of L4.5--L5.5 with 
reported H$\alpha$ pEWs, yielding a wide range of values in
$\log_{10}$(L$_{{\rm H}\alpha}$/L$_{\rm bol}$). The highest values can reach up to $-$3.7 dex, 
while the faintest upper limit lies around $-$7.5 dex. \citet{reiners08a} looked at one L4.5 and 
three L5 and placed upper limits on the $\log_{10}$(L$_{{\rm H}\alpha}$/L$_{\rm bol}$).
The sample of SDSS L5 dwarfs includes 10 L4.5, 24 L5, and 2 L5.5 sources with seven detections of
H$\alpha$ in emission, and three classified as variable \citep{schmidt07a,schmidt15}.
In their sample, \cite{pineda16} included three L5 with H$\alpha$ in emission in one of them.
They inferred a fraction of emitters of 20\% (15--35\%; 68\% confidence limit), consistent with
the interval from the SDSS sample \citep{schmidt07a,schmidt15} although to date  there are few sources.
We also note that a large number of L dwarfs show variability in their level of chromospheric
activity \citep{liebert03a,reiners08a,schmidt15,pineda16}, the most striking case being the L5.5e$+$T7 binary 
2MASSI\,J1315309$-$264951AB \citep[L5.5;][]{hall02a,hall02b,gizis02b,burgasser11b,burgasser13a,burgasser15b}.

\subsection{Radial velocity and space motion}
\label{HyadesL5:analysis_RV}

We measured the position of the H$\alpha$ line in the 1D spectrum with the {\tt{splot}}
task under IRAF\@. We identified the line at 6564.11 \AA{} with an error of 0.04 \AA{},
while the nominal air wavelength from the NIST Atomic Spectra Database is 6562.819 \AA{}.
We should add in quadrature the error on our wavelength calibration (2.3 km/s, see Section 
\ref{HyadesL5:analysis_MedRes}) and uncertainty due to the timespan in our observations (0.2 km/s).
We inferred a radial velocity of 38.0$\pm$2.9 km/s (Table \ref{tab_HyadesL5:Astro}) after 
correcting for the rotation of the Earth, the motion of the Earth around the Earth-Moon barycentre, 
the orbit of the barycentre about the Sun, and motion of the Sun relative to the specified 
standard of rest. Our value is in close agreement with the mean values derived independently 
from the ground \citep[39.1$\pm$0.2 km/s;][]{detweiler84} and from space with Hipparcos 
\citep[39.48$\pm$0.30 km/s;][]{deBruijne01}. We infer a space motion of 
[U,V,W]=[$-$42.92, $-$23.96, $-$0.72] km/s for 2M0418$+$21 (Table \ref{tab_HyadesL5:Astro}), 
in close agreement with the space motion of the cluster centre: [$-$41.1, $-$19.2, $-$1.4] km/s 
\citep{vanleeuwen09}.

According to Vizier, 2M0418$+$21 should have a Gaia $G$-band of 19.096 and should be included in the
Initial Gaia Source List \citep{smart14}, implying that precise coordinates and proper 
motions will be soon available since the second Gaia data release is expected at the end of 2017.
However, based on the equation $G-J$\,$\sim$\,0.244$\times$SpT\,$-$\,12.6332 kindly 
provided by R.\ Smart, 2M0418$+$21 might have $G$\,$\sim$\,20.4 mag, placing it at the faint
end of the Gaia catalogue. Therefore, it is unclear at this stage whether its membership
to the Hyades cluster will be settled by Gaia in the near future.

%
%%%%%%%%%%%%%%%%%%%%%%%%%%%%%%%%%%%%%%%%%%%%%%%%%%
%%%%% Table: PM, dist, Space motion %%%%%
%%%%%%%%%%%%%%%%%%%%%%%%%%%%%%%%%%%%%%%%%%%%%%%%%%
%
\begin{table}
\caption{Results of our spectroscopic analysis.}
\label{tab_HyadesL5:Astro}
\centering
\begin{tabular}{c c}
\hline
\hline
SpT & L5$\pm$0.5 \\
Spectroscopic distance & 48.8$\pm$4.0 pc \\
Radial velocity & 38.0$\pm$2.9 km/s \\
U & $-$42.92 km/s \\
V & $-$23.96 km/s \\
W & $-$0.72 km/s \\
L$_{\rm bol}$ & 2.80 $\times 10^{29}$ erg/s \\
$\log_{10}$(L$_{{\rm H}\alpha}$/L$_{\rm bol}$) & $-$6.0 dex\\
\hline
\end{tabular}
\tablefoot{We quote the spectral type, spectroscopic distance,
radial velocity, space motion, and ratio of the $\log_{10}$(L$_{{\rm H}\alpha}$/L$_{\rm bol}$) luminosity
for 2M0418$+$21\@.
}
\end{table}

\subsection{Lithium absorption}
\label{HyadesL5:analysis_Li}

The lithium test was proposed in the early 1990s to distinguish between very low-mass 
stars and brown dwarfs \citep{rebolo92,magazzu93}. The technique relies on the detection
of lithium in absorption at 6707.8 \AA{} for late-M and L dwarfs straddling the 
stellar-substellar boundary. If  lithium is detected, the object is substellar. 

\begin{figure}
\hspace{0.2cm}\resizebox{.9\hsize}{!}{
\includegraphics[angle=0]{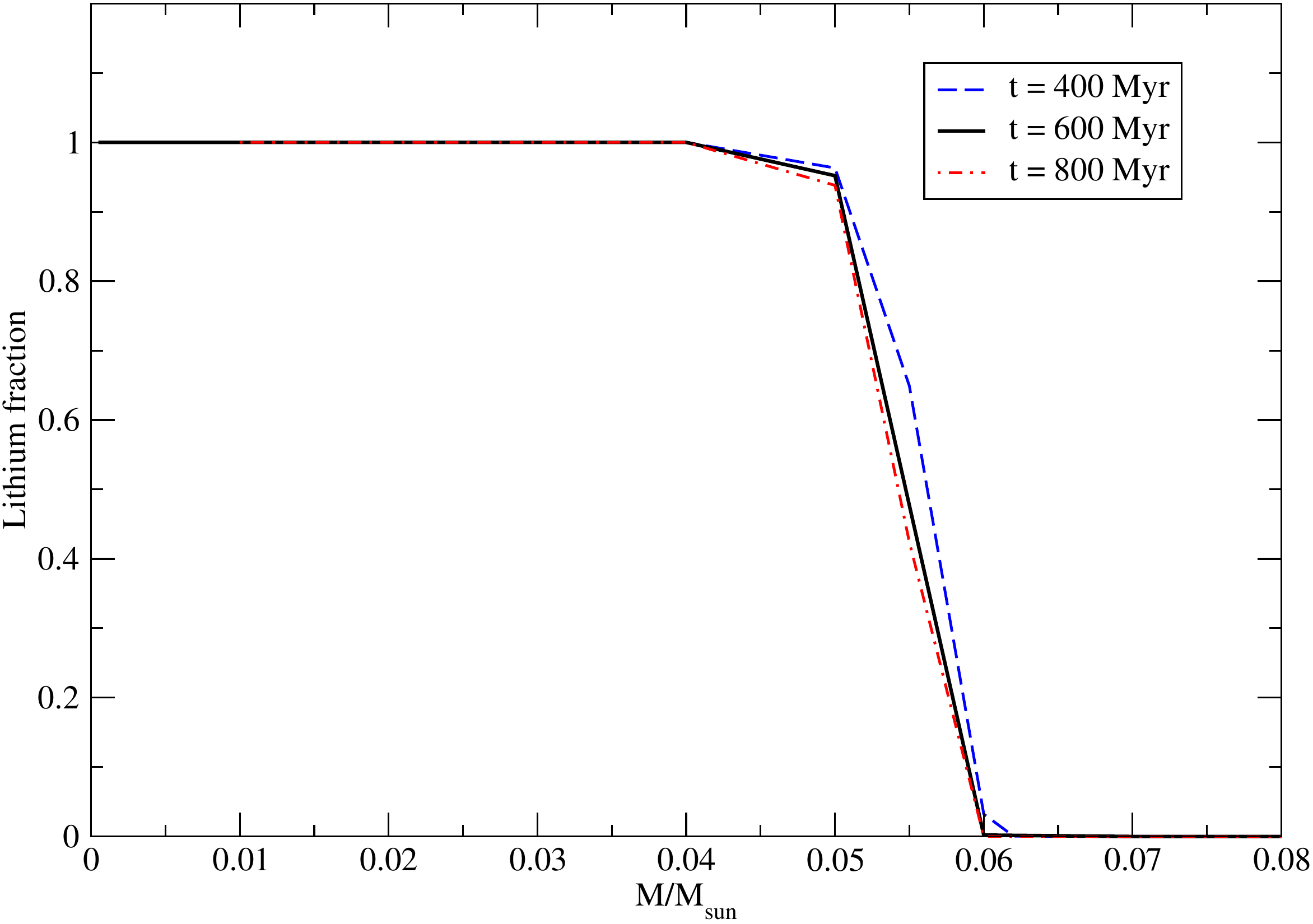}}
\caption{Fraction of lithium as a function of mass for objects below 0.08 M$_{\odot}$ at the age
of the Hyades cluster (625$\pm$50 Myr) using the AMES-Dusty model.
We added the same curves for two bracketing ages of 400 Myr and 800 Myr.
}
\label{fig_HyadesL5:M_Li}
\end{figure}

Brown dwarf evolutionary models predict a lithium conservation of 95\% for 
0.050M$_{\odot}$ brown dwarfs independent of age, whereas a 0.055 M$_{\odot}$ 
should conserve 50\% of its initial Li content and a 0.060 M$_{\odot}$ should 
have totally destroyed lithium at the age of the Hyades. In Fig. \ref{fig_HyadesL5:M_Li} 
we show the predicted lithium depletion/preservation as a function of mass at the age of the
Hyades. The observation of the lithium feature at 6707.8 \AA{} in mid-L dwarfs could 
potentially provide a critical test of consistency: models  reproduce both the 
atmospheric properties (i.e. colours) and the physics of the interior described by the
lithium content for the same mass. Many nearby old field L dwarfs show lithium in absorption 
with more than 60\% of L dwarfs with spectral types later than L5 having pseudo-equivalent
widths greater than 10--15 \AA{}, suggesting that the great majority of $>$L5 dwarfs are brown dwarfs. 

Can lithium be detected in a mid-L dwarf at the age of the Hyades? There are several 
detections of the lithium resonance doublet at 6707.8 \AA{} reported in the literature
for field L dwarfs with pEWs ranging from 1 \AA{} to 20 \AA{}
\citep{kirkpatrick00,kirkpatrick08,cruz09,zapatero14a}. For instance, \cite{faherty14a} 
and \cite{lodieu15b} have reported the detection of lithium in the two components of the 
nearest binary (L7--T0) brown dwarf to the Sun, Luhman 16AB \citep{luhman13a}. 
If preserved at the level predicted, there is a high probability that lithium can be 
detected in Hyades mid-L dwarfs. A determination of the lithium abundance in low-mass 
Hyades members would provide precise determinations of their masses and confirm them 
as genuine Hyades brown dwarfs. 

%
%%%%%%%%%%%%%%%%%%%%%%%%%%%%%%%%%%%%%%%%%%
%%%%% Figure: Lithium vs Mass  %%%%%
%%%%%%%%%%%%%%%%%%%%%%%%%%%%%%%%%%%%%%%%%%
%

\begin{figure}
  \includegraphics[width=\linewidth, angle=0]{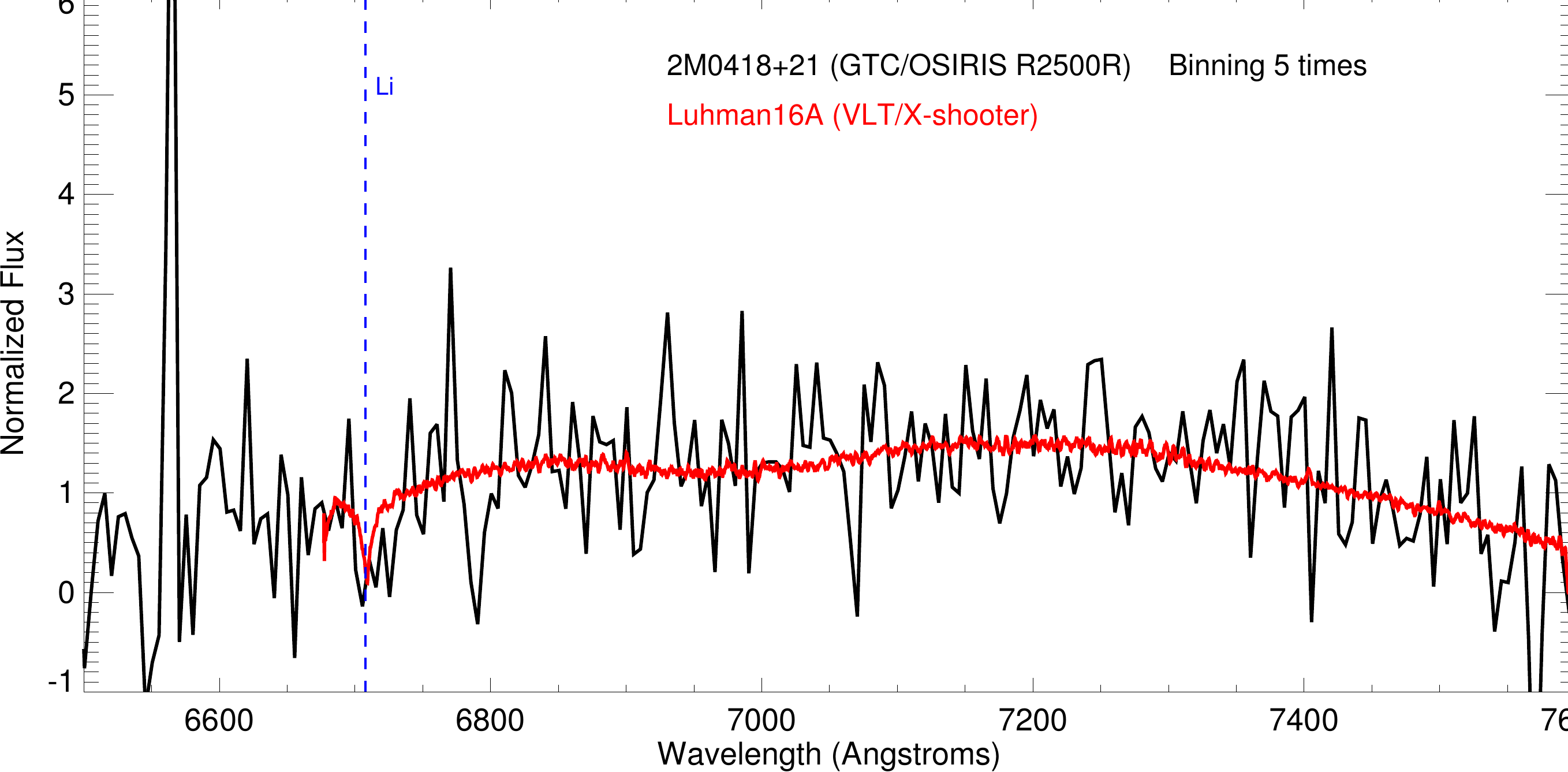}
\caption{GTC OSIRIS medium-resolution optical spectrum of 2M0415$+$21 binned by a factor of five
(black line) compared to the  smoothed (R$\sim$730) VLT X-shooter spectrum (red line)
of Luhman16A\@. The position of the lithium feature is indicated by a  blue dashed vertical line.
}
\label{fig_HyadesL5:spec_Li_compare}
\end{figure}

In Fig. \ref{fig_HyadesL5:spec_Li_compare} we compare our medium-resolution optical spectrum
of 2M0418$+$21 smoothed by a factor of nine (black line) with the unsmoothed VLT X-shooter 
spectrum (red line) of Luhman16A (L7; R$\sim$11000; \citealt{lodieu15b}). By direct 
comparison with the clear lithium absorption present in the X-shooter spectrum of Luhman\,16A,
we observe a wide feature around 6708 \AA{} that might be consistent in strength with the full 
preservation of lithium. We measured an equivalent width in the range 15--20 \AA{}.
However, we cannot claim the detection of lithium due to the low signal-to-noise of the spectrum.
If the absorption feature were confirmed as being  due to lithium with a higher quality
spectrum., it would imply a mass below 0.050 M$_{\odot}$. If true, 2M0418$+$21 would become the first 
relatively old L-type brown dwarf with  well-determined mass and age. Its photometric and 
spectroscopic properties would represent a benchmark to classify isolated objects in the field 
for which neither mass nor age can be reliably determined.

%
%%%%%%%%%%%%%%%%%%%%%%%%%%%%%%%%%%%%%%%%%%%%
%%%%% Figure: 2M0415 vs Luhman16A  %%%%%
%%%%%%%%%%%%%%%%%%%%%%%%%%%%%%%%%%%%%%%%%%%%
%

\subsection{Alkali lines}
\label{HyadesL5:Alkali_Li}
We measured the pEWs of two alkali lines present in the spectra of mid-L dwarfs, which are known to be sensitive to gravity. They are better detected than the Na{\small{I}} doublet at $\sim$8200 \AA{} owing to the low signal-to-noise of our spectra and the redness of mid-L dwarfs. On the one hand, we measured consistent pEWs of 5$\pm$2 \AA{} for the Rb{\small{I}} feature at 8946 \AA{} in both 
low-resolution spectra. We checked the strength of this line in several L dwarfs with spectral types between L3 and L8 downloaded from Sandy Leggett's homepage and in Luhman\,16AB \citep{lodieu15b}, yielding pEWs in the 3--7 \AA{} range. On the other hand, we measured a pEW of 5.0$\pm$0.5 \AA{} for the Cs{\small{I}} line at 8521 \AA{}, consistent with the 3--6 \AA{} range for L4--L6 dwarfs \citep[see Figure 17 of][]{lodieu15b}, obtained using data from \citet{kirkpatrick00} and \citet{burgasser03}. 
The fits from \cite{dupuy12}
 suggests $M_J$ = 13.40 mag for a L5 dwarf with an rms of 0.39 mag.
Our object would have $M_J$ = 13.71$\pm$0.07 mag, consistent with field dwarf within the uncertainties. At this spectral type (mid-L dwarfs), Figure 25 of \cite{liu16} suggests that young objects have fainter absolute J-band magnitude, so our target is most consistent with the field population.

%
%%%%%%%%%%%%%%%%%%%%%%%%%%%%%%%%%%%%%%%
%%%%% Conclusions %%%%%
%%%%%%%%%%%%%%%%%%%%%%%%%%%%%%%%%%%%%%%
%
\section{Conclusions}
\label{HyadesL5:conclusions}

We cross-matched the 2MASS and AllWISE databases for  identification of very  low-mass stars and brown dwarfs with proper motions consistent with membership  in the Hyades cluster. We report on the spectroscopic follow-up with GTC/OSIRIS
of one faint proper motion candidate, 2M0418$+$21, discovered in this search. According to  its photometry, proper motion/radial velocity, and  spectroscopic properties we conclude that this object is   very likely a brown dwarf member of the Hyades.   The main results obtained from  the astrometric,  photometric, and spectroscopic analysis of the 2M0418+21\@ data are as follows:
\begin{enumerate}
\item [$\bullet$]  a proper motion fully  consistent with that of the Hyades cluster, where the maximum likelihood method suggests a probability greater than 90\% for membership in the Hyades;
\item [$\bullet$] an optical spectral type of L5.0$\pm$0.5 fully consistent with the measured photometric colours;

\item [$\bullet$]  a radial velocity of 38.0$\pm$2.9 km/s, which is combined
with our proper motion to yield  a space motion fully consistent with that of higher mass Hyades members;
\item [$\bullet$] a mean spectrophotometric distance of 48.8$\pm$4.0 pc;

\item [$\bullet$] an   mass  estimated from evolutionary models and photometry  between 0.039--0.055 M$_{\odot}$, i.e. in the brown dwarf domain;

\item [$\bullet$] a spectrum  consistent with lithium in absorption at 6707.8 \AA{};
however, a higher signal-to-noise is required to confirm the full preservation of lithium  expected for  an object with mass below 0.050 M$_{\odot}$;

\item [$\bullet$] strong H$\alpha$ emission  detected in the intermediate-resolution optical spectra obtained on 19 December 2015\@ leading to  a H$\alpha$ to bolometric luminosity ratio of $-$6.0 dex  between the strongest mid-L emitters and the current upper limits set for old field L dwarfs.

\end{enumerate}

These findings make 2M0418+21\@ the faintest and oldest mid-L brown dwarf with a known age  exhibiting H$\alpha$ emission.

%
%%%%%%%%%%%%%%%%%%%%%%%%%%%%%%%
%%%%%  ACKNOWLEDGEMENTS %%%%%
%%%%%%%%%%%%%%%%%%%%%%%%%%%%%%%
%
\begin{acknowledgements}
This research has been supported by
Project No. 15345/PI/10 from the Fundaci\'on S\'eneca and the Spanish Ministry of 
Economy and Competitiveness (MINECO) under the grants AYA2015-69350-C3-2-P and 
AYA2015-69350-C3-3-P.

This work is based on observations made with the Gran Telescopio Canarias (GTC), 
operated on the island of La Palma at the Spanish Observatorio del Roque de los 
Muchachos of the Instituto de Astrof\'isica de Canarias (programme GTC37-15A led by P\'erez Garrido
and programmes GTC51--14B and GTC38-15A led by Lodieu).

This research has made use of data from the UKIDSS project defined in \citet{lawrence07}. 
UKIDSS uses the UKIRT Wide Field Camera \citep[WFCAM;][]{casali07}. The photometric system 
is described in \citet{hewett06}, and the calibration is described in \citet{hodgkin09}. 
The pipeline processing and science archive are described in Irwin et al (2009, in prep.) 
and \citet{hambly08}.

This publication makes use of data products from the Two Micron All Sky Survey
(2MASS), which is a joint project of the University of Massachusetts and
the Infrared Processing and Analysis Center/California Institute of Technology,
funded by the National Aeronautics and Space Administration and the National
Science Foundation. This publication makes use of data products from the Wide-field Infrared Survey Explorer, which is a joint project of the University of
California, Los Angeles, and the Jet Propulsion Laboratory/California Institute
of Technology, funded by the National Aeronautics and Space Administration.

Data from the Sloan Digital Sky Survey IV has been employed in the work.
Funding for the Sloan Digital Sky Survey IV has been provided by
the Alfred P. Sloan Foundation, the U.S. Department of Energy Office of
Science, and the Participating Institutions. SDSS-IV acknowledges
support and resources from the Center for High-Performance Computing at
the University of Utah. The SDSS web site is www.sdss.org.

SDSS-IV is managed by the Astrophysical Research Consortium for the 
Participating Institutions of the SDSS Collaboration including the 
Brazilian Participation Group, the Carnegie Institution for Science, 
Carnegie Mellon University, the Chilean Participation Group, the French Participation Group, Harvard-Smithsonian Center for Astrophysics, 
Instituto de Astrof\'isica de Canarias, The Johns Hopkins University, 
Kavli Institute for the Physics and Mathematics of the Universe (IPMU) / 
University of Tokyo, Lawrence Berkeley National Laboratory, 
Leibniz Institut f\"ur Astrophysik Potsdam (AIP),  
Max-Planck-Institut f\"ur Astronomie (MPIA Heidelberg), 
Max-Planck-Institut f\"ur Astrophysik (MPA Garching), 
Max-Planck-Institut f\"ur Extraterrestrische Physik (MPE), 
National Astronomical Observatory of China, New Mexico State University, 
New York University, University of Notre Dame, 
Observat\'ario Nacional / MCTI, The Ohio State University, 
Pennsylvania State University, Shanghai Astronomical Observatory, 
United Kingdom Participation Group,
Universidad Nacional Aut\'onoma de M\'exico, University of Arizona, 
University of Colorado Boulder, University of Oxford, University of Portsmouth, 
University of Utah, University of Virginia, University of Washington, University of Wisconsin, 
Vanderbilt University, and Yale University.

This publication makes use of data products from the Wide-field Infrared Survey Explorer, which is a joint project of the University of California, Los Angeles, and the Jet Propulsion Laboratory/California Institute of Technology, funded by the National Aeronautics and Space Administration.

This research has made use of the Simbad and Vizier databases, operated
at the Centre de Donn\'ees Astronomiques de Strasbourg (CDS), and
of NASA's Astrophysics Data System Bibliographic Services (ADS).
\end{acknowledgements}
%

%
%%%%%%%%%%%%%%%%%%%%%%%%%%%%%%%%%%%%%%%%%%
%%%%%%%%  Bibliography  %%%%%%%%
%%%%%%%%%%%%%%%%%%%%%%%%%%%%%%%%%%%%%%%%%%
%
%\begin{thebibliography}{}
\bibliographystyle{aa}
\bibliography{biblio} 

\begin{thebibliography}{95}
\expandafter\ifx\csname natexlab\endcsname\relax\def\natexlab#1{#1}\fi

\bibitem[{{Ahn} {et~al.}(2012{\natexlab{a}}){Ahn}, {Alexandroff}, {Allende
  Prieto}, {Anderson}, {Anderton}, {Andrews}, {Aubourg}, {Bailey}, {Balbinot},
  {Barnes}, \& et~al.}]{ahn12a}
{Ahn}, C.~P., {Alexandroff}, R., {Allende Prieto}, C., {et~al.}
  2012{\natexlab{a}}, \apjs, 203, 21

\bibitem[{{Ahn} {et~al.}(2012{\natexlab{b}}){Ahn}, {Alexandroff}, {Allende
  Prieto}, {Anderson}, {Anderton}, {Andrews}, {Aubourg}, {Bailey}, {Balbinot},
  {Barnes}, \& et~al.}]{adelman_mccarthy12}
{Ahn}, C.~P., {Alexandroff}, R., {Allende Prieto}, C., {et~al.}
  2012{\natexlab{b}}, \apjs, 203, 21

\bibitem[{{Allard} {et~al.}(2001){Allard}, {Hauschildt}, {Alexander},
  {Tamanai}, \& {Schweitzer}}]{allard01}
{Allard}, F., {Hauschildt}, P.~H., {Alexander}, D.~R., {Tamanai}, A., \&
  {Schweitzer}, A. 2001, \apj, 556, 357

\bibitem[{{Bannister} \& {Jameson}(2007)}]{bannister07}
{Bannister}, N.~P. \& {Jameson}, R.~F. 2007, \mnras, 378, L24

\bibitem[{{Baraffe} {et~al.}(1998){Baraffe}, {Chabrier}, {Allard}, \&
  {Hauschildt}}]{baraffe98}
{Baraffe}, I., {Chabrier}, G., {Allard}, F., \& {Hauschildt}, P.~H. 1998, \aap,
  337, 403

\bibitem[{{Baraffe} {et~al.}(2015){Baraffe}, {Homeier}, {Allard}, \&
  {Chabrier}}]{baraffe15}
{Baraffe}, I., {Homeier}, D., {Allard}, F., \& {Chabrier}, G. 2015, \aap, 577,
  A42

\bibitem[{{Bastian} {et~al.}(2010){Bastian}, {Covey}, \& {Meyer}}]{bastian10}
{Bastian}, N., {Covey}, K.~R., \& {Meyer}, M.~R. 2010, \araa, 48, 339

\bibitem[{{Berger} {et~al.}(2010){Berger}, {Basri}, {Fleming}, {Giampapa},
  {Gizis}, {Liebert}, {Mart{\'{\i}}n}, {Phan-Bao}, \& {Rutledge}}]{berger10}
{Berger}, E., {Basri}, G., {Fleming}, T.~A., {et~al.} 2010, \apj, 709, 332

\bibitem[{{Boesgaard} \& {Friel}(1990)}]{boesgaard90}
{Boesgaard}, A.~M. \& {Friel}, E.~D. 1990, \apj, 351, 467

\bibitem[{{Bouvier} {et~al.}(2008){Bouvier}, {Kendall}, {Meeus}, {Testi},
  {Moraux}, {Stauffer}, {James}, {Cuillandre}, {Irwin}, {McCaughrean},
  {Baraffe}, \& {Bertin}}]{bouvier08a}
{Bouvier}, J., {Kendall}, T., {Meeus}, G., {et~al.} 2008, \aap, 481, 661

\bibitem[{{Bryja} {et~al.}(1994){Bryja}, {Humphreys}, \& {Jones}}]{bryja94}
{Bryja}, C., {Humphreys}, R.~M., \& {Jones}, T.~J. 1994, \aj, 107, 246

\bibitem[{{Burgasser} {et~al.}(2003){Burgasser}, {Kirkpatrick}, {Liebert}, \&
  {Burrows}}]{burgasser03}
{Burgasser}, A.~J., {Kirkpatrick}, J.~D., {Liebert}, J., \& {Burrows}, A. 2003,
  \apj, 594, 510

\bibitem[{{Burgasser} {et~al.}(2015){Burgasser}, {Logsdon}, {Gagn{\'e}},
  {Bochanski}, {Faherty}, {West}, {Mamajek}, {Schmidt}, \&
  {Cruz}}]{burgasser15b}
{Burgasser}, A.~J., {Logsdon}, S.~E., {Gagn{\'e}}, J., {et~al.} 2015, \apjs,
  220, 18

\bibitem[{{Burgasser} {et~al.}(2013){Burgasser}, {Melis}, {Zauderer}, \&
  {Berger}}]{burgasser13a}
{Burgasser}, A.~J., {Melis}, C., {Zauderer}, B.~A., \& {Berger}, E. 2013,
  \apjl, 762, L3

\bibitem[{{Burgasser} {et~al.}(2011){Burgasser}, {Sitarski}, {Gelino},
  {Logsdon}, \& {Perrin}}]{burgasser11b}
{Burgasser}, A.~J., {Sitarski}, B.~N., {Gelino}, C.~R., {Logsdon}, S.~E., \&
  {Perrin}, M.~D. 2011, \apj, 739, 49

\bibitem[{{Burrows} \& {Liebert}(1993)}]{burrows93}
{Burrows}, A. \& {Liebert}, J. 1993, Reviews of Modern Physics, 65, 301

\bibitem[{{Casali} {et~al.}(2007){Casali}, {Adamson}, {Alves de Oliveira},
  {Almaini}, {Burch}, {Chuter}, {Elliot}, {Folger}, {Foucaud}, {Hambly},
  {Hastie}, {Henry}, {Hirst}, {Irwin}, {Ives}, {Lawrence}, {Laidlaw}, {Lee},
  {Lewis}, {Lunney}, {McLay}, {Montgomery}, {Pickup}, {Read}, {Rees}, {Robson},
  {Sekiguchi}, {Vick}, {Warren}, \& {Woodward}}]{casali07}
{Casali}, M., {Adamson}, A., {Alves de Oliveira}, C., {et~al.} 2007, \aap, 467,
  777

\bibitem[{{Casewell} {et~al.}(2008){Casewell}, {Jameson}, \&
  {Burleigh}}]{casewell08a}
{Casewell}, S.~L., {Jameson}, R.~F., \& {Burleigh}, M.~R. 2008, \mnras, 390,
  1517

\bibitem[{{Casewell} {et~al.}(2014){Casewell}, {Littlefair}, {Burleigh}, \&
  {Roy}}]{casewell14a}
{Casewell}, S.~L., {Littlefair}, S.~P., {Burleigh}, M.~R., \& {Roy}, M. 2014,
  \mnras, 441, 2644

\bibitem[{{Cepa} {et~al.}(2000){Cepa}, {Aguiar}, {Escalera},
  {Gonzalez-Serrano}, {Joven-Alvarez}, {Peraza}, {Rasilla}, {Rodriguez-Ramos},
  {Gonzalez}, {Cobos Duenas}, {Sanchez}, {Tejada}, {Bland-Hawthorn},
  {Militello}, \& {Rosa}}]{cepa00}
{Cepa}, J., {Aguiar}, M., {Escalera}, V.~G., {et~al.} 2000, in \procspie, Vol.
  4008, Optical and IR Telescope Instrumentation and Detectors, ed. M.~{Iye} \&
  A.~F. {Moorwood}, 623--631

\bibitem[{{Chabrier} \& {Baraffe}(1997)}]{chabrier97}
{Chabrier}, G. \& {Baraffe}, I. 1997, \aap, 327, 1039

\bibitem[{{Chabrier} {et~al.}(2000){Chabrier}, {Baraffe}, {Allard}, \&
  {Hauschildt}}]{chabrier00c}
{Chabrier}, G., {Baraffe}, I., {Allard}, F., \& {Hauschildt}, P. 2000, \apj,
  542, 464

\bibitem[{{Chiu} {et~al.}(2006){Chiu}, {Fan}, {Leggett}, {Golimowski}, {Zheng},
  {Geballe}, {Schneider}, \& {Brinkmann}}]{chiu06}
{Chiu}, K., {Fan}, X., {Leggett}, S.~K., {et~al.} 2006, \aj, 131, 2722

\bibitem[{{Cruz} {et~al.}(2009){Cruz}, {Kirkpatrick}, \& {Burgasser}}]{cruz09}
{Cruz}, K.~L., {Kirkpatrick}, J.~D., \& {Burgasser}, A.~J. 2009, \aj, 137, 3345

\bibitem[{{Cruz} {et~al.}(2007){Cruz}, {Reid}, {Kirkpatrick}, {Burgasser},
  {Liebert}, {Solomon}, {Schmidt}, {Allen}, {Hawley}, \& {Covey}}]{cruz07}
{Cruz}, K.~L., {Reid}, I.~N., {Kirkpatrick}, J.~D., {et~al.} 2007, \aj, 133,
  439

\bibitem[{{Cutri} {et~al.}(2003){Cutri}, {Skrutskie}, {van Dyk}, {Beichman},
  {Carpenter}, {Chester}, {Cambresy}, {Evans}, {Fowler}, {Gizis}, {Howard},
  {Huchra}, {Jarrett}, {Kopan}, {Kirkpatrick}, {Light}, {Marsh}, {McCallon},
  {Schneider}, {Stiening}, {Sykes}, {Weinberg}, {Wheaton}, {Wheelock}, \&
  {Zacarias}}]{cutri03}
{Cutri}, R.~M., {Skrutskie}, M.~F., {van Dyk}, S., {et~al.} 2003, VizieR Online
  Data Catalog, 2246, 0

\bibitem[{{Dahn} {et~al.}(2002){Dahn}, {Harris}, {Vrba}, {Guetter}, {Canzian},
  {Henden}, {Levine}, {Luginbuhl}, {Monet}, {Monet}, {Pier}, {Stone}, {Walker},
  {Burgasser}, {Gizis}, {Kirkpatrick}, {Liebert}, \& {Reid}}]{dahn02}
{Dahn}, C.~C., {Harris}, H.~C., {Vrba}, F.~J., {et~al.} 2002, \aj, 124, 1170

\bibitem[{{de Bruijne} {et~al.}(2001){de Bruijne}, {Hoogerwerf}, \& {de
  Zeeuw}}]{deBruijne01}
{de Bruijne}, J.~H.~J., {Hoogerwerf}, R., \& {de Zeeuw}, P.~T. 2001, \aap, 367,
  111

\bibitem[{{Detweiler} {et~al.}(1984){Detweiler}, {Yoss}, {Radick}, \&
  {Becker}}]{detweiler84}
{Detweiler}, H.~L., {Yoss}, K.~M., {Radick}, R.~R., \& {Becker}, S.~A. 1984,
  \aj, 89, 1038

\bibitem[{{Drilling} {et~al.}(2013){Drilling}, {Jeffery}, {Heber}, {Moehler},
  \& {Napiwotzki}}]{drilling13}
{Drilling}, J.~S., {Jeffery}, C.~S., {Heber}, U., {Moehler}, S., \&
  {Napiwotzki}, R. 2013, \aap, 551, A31

\bibitem[{{Dupuy} \& {Liu}(2012)}]{dupuy12}
{Dupuy}, T.~J. \& {Liu}, M.~C. 2012, \apjs, 201, 19

\bibitem[{{Eggen}(1998)}]{eggen98a}
{Eggen}, O.~J. 1998, \aj, 116, 284

\bibitem[{{EROS Collaboration} {et~al.}(1999){EROS Collaboration}, {Goldman},
  {Delfosse}, {Forveille}, {Afonso}, {Alard}, {Albert}, {Andersen}, {Ansari},
  {Aubourg}, {Bareyre}, {Bauer}, {Beaulieu}, {Borsenberger}, {Bouquet}, {Char},
  {Charlot}, {Couchot}, {Coutures}, {Derue}, {Ferlet}, {Fouqu{\'e}},
  {Glicenstein}, {Gould}, {Graff}, {Gros}, {Haissinski}, {Hamilton}, {Hardin},
  {de Kat}, {Kim}, {Lasserre}, {Lesquoy}, {Loup}, {Magneville}, {Mansoux},
  {Marquette}, {Mart{\'{\i}}n}, {Maurice}, {Milsztajn}, {Moniez},
  {Palanque-Delabrouille}, {Perdereau}, {Pr{\'e}vot}, {Regnault}, {Rich},
  {Spiro}, {Vidal-Madjar}, {Vigroux}, \& {Zylberajch}}]{goldman99}
{EROS Collaboration}, {Goldman}, B., {Delfosse}, X., {et~al.} 1999, \aap, 351,
  L5

\bibitem[{{Faherty} {et~al.}(2014){Faherty}, {Beletsky}, {Burgasser}, {Tinney},
  {Osip}, {Filippazzo}, \& {Simcoe}}]{faherty14a}
{Faherty}, J.~K., {Beletsky}, Y., {Burgasser}, A.~J., {et~al.} 2014, \apj, 790,
  90

\bibitem[{{Fan} {et~al.}(2000){Fan}, {Knapp}, {Strauss}, {Gunn}, {Lupton},
  {Ivezi{\'c}}, {Rockosi}, {Yanny}, {Kent}, {Schneider}, {Kirkpatrick},
  {Annis}, {Bastian}, {Berman}, {Brinkmann}, {Csabai}, {Federwitz}, {Fukugita},
  {Gurbani}, {Hennessy}, {Hindsley}, {Ichikawa}, {Lamb}, {Lindenmeyer},
  {Mantsch}, {McKay}, {Munn}, {Nash}, {Okamura}, {Pauls}, {Pier},
  {Rechenmacher}, {Rivetta}, {Sergey}, {Stoughton}, {Szalay}, {Szokoly},
  {Tucker}, {York}, \& {SDSS Collaboration}}]{fan00}
{Fan}, X., {Knapp}, G.~R., {Strauss}, M.~A., {et~al.} 2000, \aj, 119, 928

\bibitem[{{Feiden} {et~al.}(2015){Feiden}, {Jones}, \& {Chaboyer}}]{feiden15a}
{Feiden}, G.~A., {Jones}, J., \& {Chaboyer}, B. 2015, in Cambridge Workshop on
  Cool Stars, Stellar Systems, and the Sun, Vol.~18, 18th Cambridge Workshop on
  Cool Stars, Stellar Systems, and the Sun, ed. G.~T. {van Belle} \& H.~C.
  {Harris}, 171--176

\bibitem[{{Gagn{\'e}} {et~al.}(2015){Gagn{\'e}}, {Faherty}, {Cruz},
  {Lafreni{\'e}re}, {Doyon}, {Malo}, {Burgasser}, {Naud}, {Artigau},
  {Bouchard}, {Gizis}, \& {Albert}}]{gagne15c}
{Gagn{\'e}}, J., {Faherty}, J.~K., {Cruz}, K.~L., {et~al.} 2015, \apjs, 219, 33

\bibitem[{{Geballe} {et~al.}(2002){Geballe}, {Knapp}, {Leggett}, {Fan},
  {Golimowski}, {Anderson}, {Brinkmann}, {Csabai}, {Gunn}, {Hawley},
  {Hennessy}, {Henry}, {Hill}, {Hindsley}, {Ivezi{\'c}}, {Lupton}, {McDaniel},
  {Munn}, {Narayanan}, {Peng}, {Pier}, {Rockosi}, {Schneider}, {Smith},
  {Strauss}, {Tsvetanov}, {Uomoto}, {York}, \& {Zheng}}]{geballe02}
{Geballe}, T.~R., {Knapp}, G.~R., {Leggett}, S.~K., {et~al.} 2002, \apj, 564,
  466

\bibitem[{{Gebran} {et~al.}(2010){Gebran}, {Vick}, {Monier}, \&
  {Fossati}}]{gebran10}
{Gebran}, M., {Vick}, M., {Monier}, R., \& {Fossati}, L. 2010, \aap, 523, A71

\bibitem[{{Gizis}(2002)}]{gizis02b}
{Gizis}, J.~E. 2002, \apj, 575, 484

\bibitem[{{Goldman} {et~al.}(2013){Goldman}, {R{\"o}ser}, {Schilbach},
  {Magnier}, {Olczak}, {Henning}, {Juri{\'c}}, {Schlafly}, {Chen}, {Platais},
  {Burgett}, {Hodapp}, {Heasley}, {Kudritzki}, {Morgan}, {Price}, {Tonry}, \&
  {Wainscoat}}]{goldman13}
{Goldman}, B., {R{\"o}ser}, S., {Schilbach}, E., {et~al.} 2013, \aap, 559, A43

\bibitem[{{Golimowski} {et~al.}(2004){Golimowski}, {Leggett}, {Marley}, {Fan},
  {Geballe}, {Knapp}, {Vrba}, {Henden}, {Luginbuhl}, {Guetter}, {Munn},
  {Canzian}, {Zheng}, {Tsvetanov}, {Chiu}, {Glazebrook}, {Hoversten},
  {Schneider}, \& {Brinkmann}}]{golimowski04a}
{Golimowski}, D.~A., {Leggett}, S.~K., {Marley}, M.~S., {et~al.} 2004, \aj,
  127, 3516

\bibitem[{{Hall}(2002{\natexlab{a}})}]{hall02a}
{Hall}, P.~B. 2002{\natexlab{a}}, \apjl, 580, L77

\bibitem[{{Hall}(2002{\natexlab{b}})}]{hall02b}
{Hall}, P.~B. 2002{\natexlab{b}}, \apjl, 580, L77

\bibitem[{{Hambly} {et~al.}(2008){Hambly}, {Collins}, {Cross}, {Mann}, {Read},
  {Sutorius}, {Bond}, {Bryant}, {Emerson}, {Lawrence}, {Rimoldini}, {Stewart},
  {Williams}, {Adamson}, {Hirst}, {Dye}, \& {Warren}}]{hambly08}
{Hambly}, N.~C., {Collins}, R.~S., {Cross}, N.~J.~G., {et~al.} 2008, \mnras,
  384, 637

\bibitem[{{Harrington} \& {Dahn}(1980)}]{harrington80}
{Harrington}, R.~S. \& {Dahn}, C.~C. 1980, \aj, 85, 454

\bibitem[{{Henry} {et~al.}(2006){Henry}, {Jao}, {Subasavage}, {Beaulieu},
  {Ianna}, {Costa}, \& {M{\'e}ndez}}]{henry06}
{Henry}, T.~J., {Jao}, W.-C., {Subasavage}, J.~P., {et~al.} 2006, \aj, 132,
  2360

\bibitem[{{Hewett} {et~al.}(2006){Hewett}, {Warren}, {Leggett}, \&
  {Hodgkin}}]{hewett06}
{Hewett}, P.~C., {Warren}, S.~J., {Leggett}, S.~K., \& {Hodgkin}, S.~T. 2006,
  \mnras, 367, 454

\bibitem[{{Hodgkin} {et~al.}(2009){Hodgkin}, {Irwin}, {Hewett}, \&
  {Warren}}]{hodgkin09}
{Hodgkin}, S.~T., {Irwin}, M.~J., {Hewett}, P.~C., \& {Warren}, S.~J. 2009,
  \mnras, 394, 675

\bibitem[{{H{\o}g} {et~al.}(2000){H{\o}g}, {Fabricius}, {Makarov}, {Urban},
  {Corbin}, {Wycoff}, {Bastian}, {Schwekendiek}, \& {Wicenec}}]{hog00}
{H{\o}g}, E., {Fabricius}, C., {Makarov}, V.~V., {et~al.} 2000, \aap, 355, L27

\bibitem[{{Hogan} {et~al.}(2008){Hogan}, {Jameson}, {Casewell}, {Osbourne}, \&
  {Hambly}}]{hogan08}
{Hogan}, E., {Jameson}, R.~F., {Casewell}, S.~L., {Osbourne}, S.~L., \&
  {Hambly}, N.~C. 2008, \mnras, 388, 495

\bibitem[{{Kirkpatrick} {et~al.}(2008){Kirkpatrick}, {Cruz}, {Barman},
  {Burgasser}, {Looper}, {Tinney}, {Gelino}, {Lowrance}, {Liebert},
  {Carpenter}, {Hillenbrand}, \& {Stauffer}}]{kirkpatrick08}
{Kirkpatrick}, J.~D., {Cruz}, K.~L., {Barman}, T.~S., {et~al.} 2008, \apj, 689,
  1295

\bibitem[{{Kirkpatrick} {et~al.}(2001){Kirkpatrick}, {Dahn}, {Monet}, {Reid},
  {Gizis}, {Liebert}, \& {Burgasser}}]{kirkpatrick01}
{Kirkpatrick}, J.~D., {Dahn}, C.~C., {Monet}, D.~G., {et~al.} 2001, \aj, 121,
  3235

\bibitem[{{Kirkpatrick} {et~al.}(1999){Kirkpatrick}, {Reid}, {Liebert},
  {Cutri}, {Nelson}, {Beichman}, {Dahn}, {Monet}, {Gizis}, \&
  {Skrutskie}}]{kirkpatrick99}
{Kirkpatrick}, J.~D., {Reid}, I.~N., {Liebert}, J., {et~al.} 1999, \apj, 519,
  802

\bibitem[{{Kirkpatrick} {et~al.}(2000){Kirkpatrick}, {Reid}, {Liebert},
  {Gizis}, {Burgasser}, {Monet}, {Dahn}, {Nelson}, \&
  {Williams}}]{kirkpatrick00}
{Kirkpatrick}, J.~D., {Reid}, I.~N., {Liebert}, J., {et~al.} 2000, \aj, 120,
  447

\bibitem[{{Knapp} {et~al.}(2004){Knapp}, {Leggett}, {Fan}, {Marley}, {Geballe},
  {Golimowski}, {Finkbeiner}, {Gunn}, {Hennawi}, {Ivezi{\'c}}, {Lupton},
  {Schlegel}, {Strauss}, {Tsvetanov}, {Chiu}, {Hoversten}, {Glazebrook},
  {Zheng}, {Hendrickson}, {Williams}, {Uomoto}, {Vrba}, {Henden}, {Luginbuhl},
  {Guetter}, {Munn}, {Canzian}, {Schneider}, \& {Brinkmann}}]{knapp04}
{Knapp}, G.~R., {Leggett}, S.~K., {Fan}, X., {et~al.} 2004, \aj, 127, 3553

\bibitem[{{Landolt} \& {Uomoto}(2007)}]{landolt07}
{Landolt}, A.~U. \& {Uomoto}, A.~K. 2007, \aj, 133, 768

\bibitem[{{Lawrence} {et~al.}(2007){Lawrence}, {Warren}, {Almaini}, {Edge},
  {Hambly}, {Jameson}, {Lucas}, {Casali}, {Adamson}, {Dye}, {Emerson},
  {Foucaud}, {Hewett}, {Hirst}, {Hodgkin}, {Irwin}, {Lodieu}, {McMahon},
  {Simpson}, {Smail}, {Mortlock}, \& {Folger}}]{lawrence07}
{Lawrence}, A., {Warren}, S.~J., {Almaini}, O., {et~al.} 2007, \mnras, 379,
  1599

\bibitem[{{Leggett} {et~al.}(2002){Leggett}, {Golimowski}, {Fan}, {Geballe},
  {Knapp}, {Brinkmann}, {Csabai}, {Gunn}, {Hawley}, {Henry}, {Hindsley},
  {Ivezi{\'c}}, {Lupton}, {Pier}, {Schneider}, {Smith}, {Strauss}, {Uomoto}, \&
  {York}}]{leggett02}
{Leggett}, S.~K., {Golimowski}, D.~A., {Fan}, X., {et~al.} 2002, \apj, 564, 452

\bibitem[{{L{\'e}pine} \& {Shara}(2005)}]{lepine05d}
{L{\'e}pine}, S. \& {Shara}, M.~M. 2005, \aj, 129, 1483

\bibitem[{{Liebert} {et~al.}(2003){Liebert}, {Kirkpatrick}, {Cruz}, {Reid},
  {Burgasser}, {Tinney}, \& {Gizis}}]{liebert03a}
{Liebert}, J., {Kirkpatrick}, J.~D., {Cruz}, K.~L., {et~al.} 2003, \aj, 125,
  343

\bibitem[{{Liu} {et~al.}(2016){Liu}, {Dupuy}, \& {Allers}}]{liu16}
{Liu}, M.~C., {Dupuy}, T.~J., \& {Allers}, K.~N. 2016, \apj, 833, 96

\bibitem[{{Lodieu} {et~al.}(2014){Lodieu}, {Boudreault}, \&
  {B{\'e}jar}}]{lodieu14b}
{Lodieu}, N., {Boudreault}, S., \& {B{\'e}jar}, V.~J.~S. 2014, \mnras, 445,
  3908

\bibitem[{{Lodieu} {et~al.}(2007){Lodieu}, {Hambly}, {Jameson}, {Hodgkin},
  {Carraro}, \& {Kendall}}]{lodieu07a}
{Lodieu}, N., {Hambly}, N.~C., {Jameson}, R.~F., {et~al.} 2007, \mnras, 374,
  372

\bibitem[{{Lodieu} {et~al.}(2015){Lodieu}, {Zapatero Osorio}, {Rebolo},
  {B{\'e}jar}, {Pavlenko}, \& {P{\'e}rez-Garrido}}]{lodieu15b}
{Lodieu}, N., {Zapatero Osorio}, M.~R., {Rebolo}, R., {et~al.} 2015, \aap, 581,
  A73

\bibitem[{{Lucas} {et~al.}(2008){Lucas}, {Hoare}, {Longmore}, {Schr{\"o}der},
  {Davis}, {Adamson}, {Bandyopadhyay}, {de Grijs}, {Smith}, {Gosling},
  {Mitchison}, {G{\'a}sp{\'a}r}, {Coe}, {Tamura}, {Parker}, {Irwin}, {Hambly},
  {Bryant}, {Collins}, {Cross}, {Evans}, {Gonzalez-Solares}, {Hodgkin},
  {Lewis}, {Read}, {Riello}, {Sutorius}, {Lawrence}, {Drew}, {Dye}, \&
  {Thompson}}]{lucas08}
{Lucas}, P.~W., {Hoare}, M.~G., {Longmore}, A., {et~al.} 2008, \mnras, 391, 136

\bibitem[{{Luhman}(2012)}]{luhman12b}
{Luhman}, K.~L. 2012, \araa, 50, 65

\bibitem[{{Luhman}(2013)}]{luhman13a}
{Luhman}, K.~L. 2013, \apjl, 767, L1

\bibitem[{{Madsen} {et~al.}(2002){Madsen}, {Dravins}, \&
  {Lindegren}}]{madsen02}
{Madsen}, S., {Dravins}, D., \& {Lindegren}, L. 2002, \aap, 381, 446

\bibitem[{{Magazzu} {et~al.}(1993){Magazzu}, {Martin}, \& {Rebolo}}]{magazzu93}
{Magazzu}, A., {Martin}, E.~L., \& {Rebolo}, R. 1993, \apjl, 404, L17

\bibitem[{{Mermilliod}(1981)}]{mermilliod81}
{Mermilliod}, J.~C. 1981, \aap, 97, 235

\bibitem[{{Monet} {et~al.}(2003){Monet}, {Levine}, {Canzian}, {Ables}, {Bird},
  {Dahn}, {Guetter}, {Harris}, {Henden}, {Leggett}, {Levison}, {Luginbuhl},
  {Martini}, {Monet}, {Munn}, {Pier}, {Rhodes}, {Riepe}, {Sell}, {Stone},
  {Vrba}, {Walker}, {Westerhout}, {Brucato}, {Reid}, {Schoening}, {Hartley},
  {Read}, \& {Tritton}}]{monet03}
{Monet}, D.~G., {Levine}, S.~E., {Canzian}, B., {et~al.} 2003, \aj, 125, 984

\bibitem[{{Perryman} {et~al.}(1998){Perryman}, {Brown}, {Lebreton}, {Gomez},
  {Turon}, {Cayrel de Strobel}, {Mermilliod}, {Robichon}, {Kovalevsky}, \&
  {Crifo}}]{perryman98}
{Perryman}, M.~A.~C., {Brown}, A.~G.~A., {Lebreton}, Y., {et~al.} 1998, \aap,
  331, 81

\bibitem[{{Pineda} {et~al.}(2016){Pineda}, {Hallinan}, {Kirkpatrick}, {Cotter},
  {Kao}, \& {Mooley}}]{pineda16}
{Pineda}, J.~S., {Hallinan}, G., {Kirkpatrick}, J.~D., {et~al.} 2016, ArXiv
  e-prints [\eprint[arXiv]{1604.03941}]

\bibitem[{{Rebolo} {et~al.}(1992){Rebolo}, {Martin}, \& {Magazzu}}]{rebolo92}
{Rebolo}, R., {Martin}, E.~L., \& {Magazzu}, A. 1992, \apjl, 389, L83

\bibitem[{{Reid} {et~al.}(2000){Reid}, {Kirkpatrick}, {Gizis}, {Dahn}, {Monet},
  {Williams}, {Liebert}, \& {Burgasser}}]{reid00}
{Reid}, I.~N., {Kirkpatrick}, J.~D., {Gizis}, J.~E., {et~al.} 2000, \aj, 119,
  369

\bibitem[{{Reiners} \& {Basri}(2008)}]{reiners08a}
{Reiners}, A. \& {Basri}, G. 2008, \apj, 684, 1390

\bibitem[{{Sanders}(1971)}]{sanders71}
{Sanders}, W.~L. 1971, \aap, 14, 226

\bibitem[{{Schmidt} {et~al.}(2007){Schmidt}, {Cruz}, {Bongiorno}, {Liebert}, \&
  {Reid}}]{schmidt07a}
{Schmidt}, S.~J., {Cruz}, K.~L., {Bongiorno}, B.~J., {Liebert}, J., \& {Reid},
  I.~N. 2007, \aj, 133, 2258

\bibitem[{{Schmidt} {et~al.}(2015){Schmidt}, {Hawley}, {West}, {Bochanski},
  {Davenport}, {Ge}, \& {Schneider}}]{schmidt15}
{Schmidt}, S.~J., {Hawley}, S.~L., {West}, A.~A., {et~al.} 2015, \aj, 149, 158

\bibitem[{{Schmidt} {et~al.}(2014){Schmidt}, {West}, {Bochanski}, {Hawley}, \&
  {Kielty}}]{schmidt14a}
{Schmidt}, S.~J., {West}, A.~A., {Bochanski}, J.~J., {Hawley}, S.~L., \&
  {Kielty}, C. 2014, \pasp, 126, 642

\bibitem[{{Schmidt} {et~al.}(2010){Schmidt}, {West}, {Hawley}, \&
  {Pineda}}]{schmidt10b}
{Schmidt}, S.~J., {West}, A.~A., {Hawley}, S.~L., \& {Pineda}, J.~S. 2010, \aj,
  139, 1808

\bibitem[{{Siess} {et~al.}(2000){Siess}, {Dufour}, \& {Forestini}}]{siess00}
{Siess}, L., {Dufour}, E., \& {Forestini}, M. 2000, \aap, 358, 593

\bibitem[{{Sion} {et~al.}(2009){Sion}, {Holberg}, {Oswalt}, {McCook}, \&
  {Wasatonic}}]{sion09}
{Sion}, E.~M., {Holberg}, J.~B., {Oswalt}, T.~D., {McCook}, G.~P., \&
  {Wasatonic}, R. 2009, \aj, 138, 1681

\bibitem[{{Skrutskie} {et~al.}(2006){Skrutskie}, {Cutri}, {Stiening},
  {Weinberg}, {Schneider}, {Carpenter}, {Beichman}, {Capps}, {Chester},
  {Elias}, {Huchra}, {Liebert}, {Lonsdale}, {Monet}, {Price}, {Seitzer},
  {Jarrett}, {Kirkpatrick}, {Gizis}, {Howard}, {Evans}, {Fowler}, {Fullmer},
  {Hurt}, {Light}, {Kopan}, {Marsh}, {McCallon}, {Tam}, {Van Dyk}, \&
  {Wheelock}}]{skrutskie06}
{Skrutskie}, M.~F., {Cutri}, R.~M., {Stiening}, R., {et~al.} 2006, \aj, 131,
  1163

\bibitem[{{Smart} \& {Nicastro}(2014)}]{smart14}
{Smart}, R.~L. \& {Nicastro}, L. 2014, \aap, 570, A87

\bibitem[{{Tody}(1986)}]{tody86}
{Tody}, D. 1986, in \procspie, Vol. 627, Instrumentation in astronomy VI, ed.
  D.~L. {Crawford}, 733

\bibitem[{{Tody}(1993)}]{tody93}
{Tody}, D. 1993, in Astronomical Society of the Pacific Conference Series,
  Vol.~52, Astronomical Data Analysis Software and Systems II, ed. R.~J.
  {Hanisch}, R.~J.~V. {Brissenden}, \& J.~{Barnes}, 173

\bibitem[{{van Leeuwen}(2007)}]{vanleeuwen07}
{van Leeuwen}, F. 2007, \aap, 474, 653

\bibitem[{{van Leeuwen}(2009)}]{vanleeuwen09}
{van Leeuwen}, F. 2009, A\&A, 497, 209

\bibitem[{{Vrba} {et~al.}(2004){Vrba}, {Henden}, {Luginbuhl}, {Guetter},
  {Munn}, {Canzian}, {Burgasser}, {Kirkpatrick}, {Fan}, {Geballe},
  {Golimowski}, {Knapp}, {Leggett}, {Schneider}, \& {Brinkmann}}]{vrba04}
{Vrba}, F.~J., {Henden}, A.~A., {Luginbuhl}, C.~B., {et~al.} 2004, \aj, 127,
  2948

\bibitem[{{Wright} {et~al.}(2010){Wright}, {Eisenhardt}, {Mainzer}, {Ressler},
  {Cutri}, {Jarrett}, {Kirkpatrick}, {Padgett}, {McMillan}, {Skrutskie},
  {Stanford}, {Cohen}, {Walker}, {Mather}, {Leisawitz}, {Gautier}, {McLean},
  {Benford}, {Lonsdale}, {Blain}, {Mendez}, {Irace}, {Duval}, {Liu}, {Royer},
  {Heinrichsen}, {Howard}, {Shannon}, {Kendall}, {Walsh}, {Larsen}, {Cardon},
  {Schick}, {Schwalm}, {Abid}, {Fabinsky}, {Naes}, \& {Tsai}}]{wright10}
{Wright}, E.~L., {Eisenhardt}, P.~R.~M., {Mainzer}, A.~K., {et~al.} 2010, \aj,
  140, 1868

\bibitem[{{York} {et~al.}(2000){York}, {Adelman}, {Anderson}, {Anderson},
  {Annis}, {Bahcall}, {Bakken}, {Barkhouser}, {Bastian}, {Berman}, {Boroski},
  {Bracker}, {Briegel}, {Briggs}, {Brinkmann}, {Brunner}, {Burles}, {Carey},
  {Carr}, {Castander}, {Chen}, {Colestock}, {Connolly}, {Crocker}, {Csabai},
  {Czarapata}, {Davis}, {Doi}, {Dombeck}, {Eisenstein}, {Ellman}, {Elms},
  {Evans}, {Fan}, {Federwitz}, {Fiscelli}, {Friedman}, {Frieman}, {Fukugita},
  {Gillespie}, {Gunn}, {Gurbani}, {de Haas}, {Haldeman}, {Harris}, {Hayes},
  {Heckman}, {Hennessy}, {Hindsley}, {Holm}, {Holmgren}, {Huang}, {Hull},
  {Husby}, {Ichikawa}, {Ichikawa}, {Ivezi{\'c}}, {Kent}, {Kim}, {Kinney},
  {Klaene}, {Kleinman}, {Kleinman}, {Knapp}, {Korienek}, {Kron}, {Kunszt},
  {Lamb}, {Lee}, {Leger}, {Limmongkol}, {Lindenmeyer}, {Long}, {Loomis},
  {Loveday}, {Lucinio}, {Lupton}, {MacKinnon}, {Mannery}, {Mantsch}, {Margon},
  {McGehee}, {McKay}, {Meiksin}, {Merelli}, {Monet}, {Munn}, {Narayanan},
  {Nash}, {Neilsen}, {Neswold}, {Newberg}, {Nichol}, {Nicinski}, {Nonino},
  {Okada}, {Okamura}, {Ostriker}, {Owen}, {Pauls}, {Peoples}, {Peterson},
  {Petravick}, {Pier}, {Pope}, {Pordes}, {Prosapio}, {Rechenmacher}, {Quinn},
  {Richards}, {Richmond}, {Rivetta}, {Rockosi}, {Ruthmansdorfer}, {Sandford},
  {Schlegel}, {Schneider}, {Sekiguchi}, {Sergey}, {Shimasaku}, {Siegmund},
  {Smee}, {Smith}, {Snedden}, {Stone}, {Stoughton}, {Strauss}, {Stubbs},
  {SubbaRao}, {Szalay}, {Szapudi}, {Szokoly}, {Thakar}, {Tremonti}, {Tucker},
  {Uomoto}, {Vanden Berk}, {Vogeley}, {Waddell}, {Wang}, {Watanabe},
  {Weinberg}, {Yanny}, {Yasuda}, \& {SDSS Collaboration}}]{york00}
{York}, D.~G., {Adelman}, J., {Anderson}, Jr., J.~E., {et~al.} 2000, \aj, 120,
  1579

\bibitem[{{Zapatero Osorio} {et~al.}(2014){Zapatero Osorio}, {B{\'e}jar},
  {Miles-P{\'a}ez}, {Pe{\~n}a Ram{\'{\i}}rez}, {Rebolo}, \&
  {Pall{\'e}}}]{zapatero14a}
{Zapatero Osorio}, M.~R., {B{\'e}jar}, V.~J.~S., {Miles-P{\'a}ez}, P.~A.,
  {et~al.} 2014, \aap, 568, A6

\bibitem[{{Zapatero Osorio} {et~al.}(2004){Zapatero Osorio}, {Lane},
  {Pavlenko}, {Mart{\'{\i}}n}, {Britton}, \& {Kulkarni}}]{zapatero04b}
{Zapatero Osorio}, M.~R., {Lane}, B.~F., {Pavlenko}, Y., {et~al.} 2004, \apj,
  615, 958

\end{thebibliography}

%
%\end{thebibliography}
%

\end{document}